\documentclass[floats, floatfix, showpacs, amssymb, prd, twocolumn, superscriptaddress, nofootinbib, nolongbibliography, reprint]{revtex4-2}

\usepackage{amssymb,amsmath,verbatim,mathtools,needspace,enumitem,etoolbox,graphicx,physics,microtype,afterpage}
\usepackage{gensymb}

\usepackage[dvipsnames, usenames]{xcolor}

\definecolor{linkcolor}{rgb}{0.0,0.3,0.5}
\usepackage[unicode, colorlinks=true, linkcolor=linkcolor, citecolor=linkcolor, filecolor=linkcolor, urlcolor=linkcolor, pdfusetitle]{hyperref}
\usepackage{textcomp}
\usepackage[all]{hypcap}
\usepackage[T1]{fontenc}
\usepackage[utf8]{inputenc}
\usepackage{tabularx}
\usepackage[normalem]{ulem}
\usepackage[thinc]{esdiff}
\usepackage{graphicx,epsfig,epsf}
\usepackage{lipsum}
\usepackage{float}
\usepackage{cleveref}

\renewcommand{\vec}[1]{\mathbf{#1}}

\def\aj{\rm{Astron.~J.}}

\def\apj{\rm{Astrophys.~J.}}
\def\apjl{\rm{Astrophys.~J.~ Lett.}}
\def\apjs{\rm{ApJS}}
\def\aap{\rm{Astron.~Astrophys.}}

\def\prd{\rm{Phys.~Rev.~D}}

\newcommand{\dallas}{\affiliation{Department of Physics, The University of Texas at Dallas, Richardson, Texas 75080, USA}}

\newcommand{\israel}{\affiliation{Department of Particle Physics \& Astrophysics, Weizmann Institute of Science, Rehovot 7610001, Israel}}

\newcommand{\umani}{\affiliation{Department of Physics and Astronomy \& Winnipeg Institute for Theoretical Physics, University of Manitoba, Winnipeg, R3T 2N2, Canada}}

\newcommand\orcid[1]{\href{https://orcid.org/#1}{$\!$\includegraphics[scale=0.006]{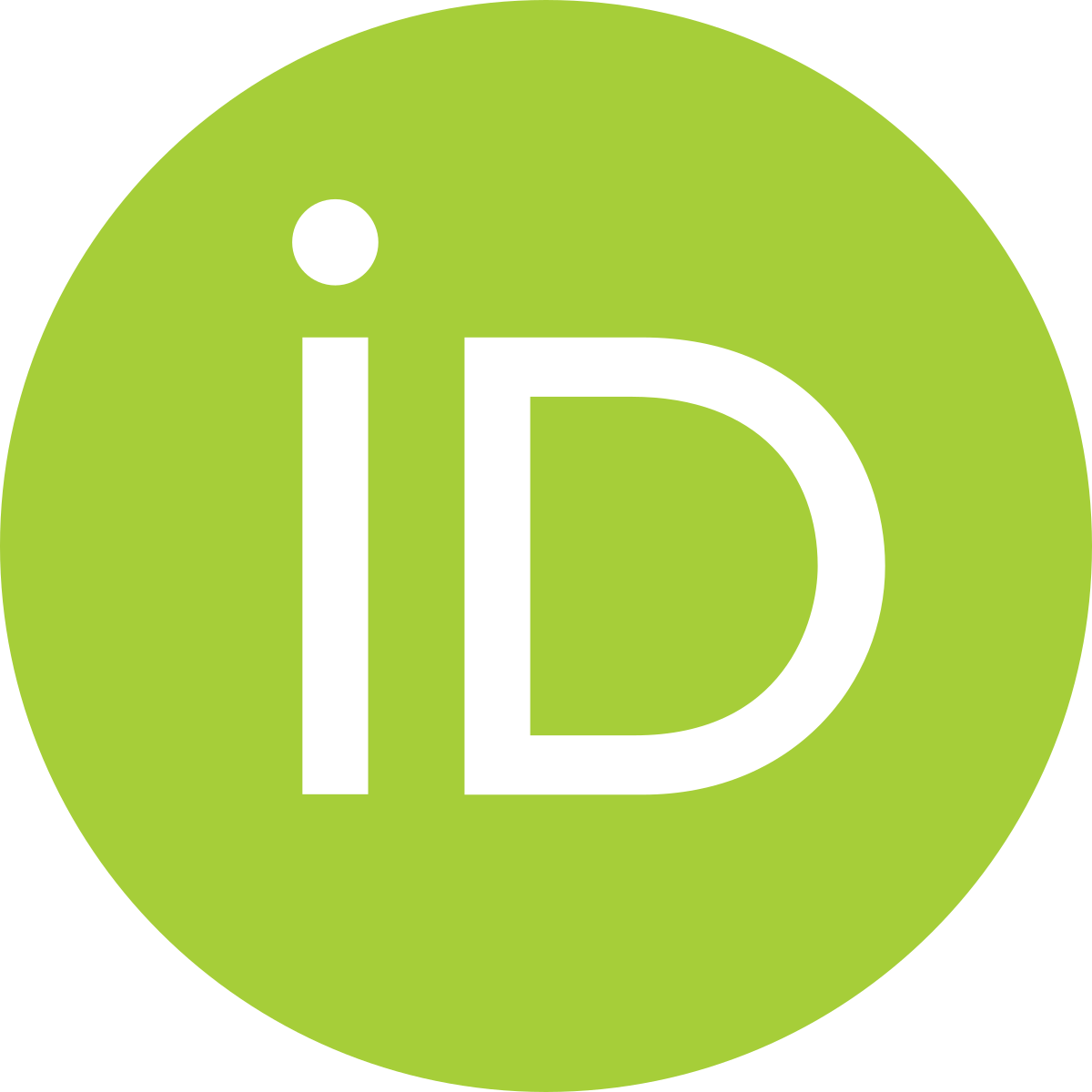} $\!\!$}}

\begin{document}

\title{
Detecting regular precession using a new gravitational waveform model directly parameterized by both precession amplitude and frequency
}

\author{Tamanjyot Singh \orcid{0009-0006-4040-4407}}
\email{tamanjyot@utdallas.edu}
\dallas

\author{Evangelos Stoikos \orcid{0000-0002-1043-3673}}
\dallas

\author{Saif Ali \orcid{0000-0002-6971-4971}}
\email{saif.ali@weizmann.ac.il}
\israel

\author{Nathan Steinle \orcid{0000-0003-0658-402X}}
\email{nathan.steinle@umanitoba.ca}
\umani

\author{Michael Kesden \orcid{0000-0002-5987-1471}}
\email{kesden@utdallas.edu}
\dallas

\author{Lindsay King}
\email{Lindsay.King@utdallas.edu}
\dallas

\date{\today}

\begin{abstract}

Nearly 210 binary black hole (BBH) mergers have been observed as gravitational-wave (GW) signals by the LIGO-Virgo-KAGRA network during its first four observing runs. Generically, both BBHs are spinning and their spins are misaligned with the orbital angular momentum $\Vec{L}$ of the binary and each other. These misaligned spins cause $\vec{L}$ to precess in a cone with dimensionless precession amplitude $\tilde{\theta}$ and frequency $\tilde{\Omega}$ about the nearly constant direction of the total angular momentum.  This precession modulates the amplitude and phase of the observed GWs. We propose a geometrical model of regularly precessing (RP) waveforms based on the hierarchy of the orbital, precessional, and radiation-reaction timescales in coalescing binaries that incorporate $\tilde{\theta}$ and $\tilde{\Omega}$ directly as parameters.  We investigate how these waveforms vary as functions of these precessional parameters, as well as binary orientation and sky location. We use the Lindblom criterion to estimate that precession can be detected in a RP source with signal-to-noise ratio $\rho$ when the mismatch $\epsilon$ with a non-precessing (NP) source with otherwise identical parameters exceeds $1/2\rho^2$. Precession is most detectable when $\Vec{L}$ precesses through configurations we call +~nulls during the inspiral.  At binary orientations and sky locations corresponding to +~nulls, a NP source emits pure +-polarization and yet the GW detector is insensitive to this polarization.  The large mismatch between a RP source and this vanishing NP signal enhances the detectability of precession according to the Lindblom criterion.  We also explore the detectability of precession as a function of redshift $z$ for different BBH populations.  We find that for BBHs with isotropically oriented maximal spins, precession is detectable in a majority of systems out to $z \approx 0.3$ for chirp masses $10 \lesssim \mathcal{M}/M_\odot \lesssim 40$ and mass ratios $q \gtrsim 0.5$.  Reduced spin magnitudes or greater alignment between the spins and $\Vec{L}$ greatly reduce the detectability of precession, making it difficult to observe beyond $z \approx 0.1$.
\end{abstract}

\maketitle

\section{Introduction}
\label{sec: Intro}

One of the main predictions of general relativity is the emission of GWs during compact binary coalescences (CBCs).  After almost a century of attempts, the first successful observation of GWs from a binary black-hole (BBH) merger was reported from the LIGO and the VIRGO collaboration in 2016~\cite{2016}. Advanced LIGO (aLIGO)~\cite{aLIGO} and advanced VIRGO (aVIRGO)~\cite{aVIRGO} performed three more runs after the first successful one, and brought the total number of GW detections to 90 events~\cite{GWTC3} at the end of the third observing run. 
Out of these 90 events, 84 are confirmed to be from BBH mergers, four of them are from neutron star black hole mergers, and the last remaining two were potentially produced from a binary neutron star merger. An additional 128 candidates have been reported in the first part of the ongoing fourth observing run (O4) \cite{LIGOO4PublicAlerts,LIGOGWTC4Aug}, with most of the signals originating from merging binary black holes.

Recently, the Kamioka Gravitational Wave Detector (KAGRA)~\cite{KAGRA1, KAGRA2, KAGRA3} has become operational and will join  aLIGO and aVIRGO to form the LVK detection network for future observational runs. Hundreds (millions) of additional events are expected in the next years (decades) due to the continuous improvements in the sensitivity of the LVK network, together with the proposed next generation of detectors, such as the Einstein Telescope~\cite{ET}, the Cosmic Explorer ~\cite{CE}, the Deci-Hertz Interferometer Gravitational Wave Observatory ~\cite{DECIGO} and the Laser Interferometer Space Antenna ~\cite{LISA}.

In the O3b and O4a (offline search) runs, the LVK collaboration used four pipelines to search for GW signals. Out of the four, three of them, GstLAL~\cite{Gst1, Gst2, Gst3}, Multi-band template analysis~\cite{MBTA1, MBTA2} and PyCBC~\cite{Usman_2016,Nitz_2017} perform their searches by using already constructed CBC waveform templates, while the remaining method, cWB~\cite{cWb1, cWB2} used phenomenological waveforms that made more minimal assumptions about the behavior of the sources. The CBC search pipelines utilize match-filter analysis~\cite{Wainstein1970,Owen_Satyaprakash1999} to compare the signal received by the ground-based detectors with a bank of carefully constructed templates.  This allows the extraction of information embedded in the signal about the source's properties. 
The more strongly these templates depend on the individual masses and spins of the BHs, the more precisely these source properties can be constrained. Although the masses and some aligned combinations of the spins are well constrained, the short duration of the signals as well as the subtle effects of the misaligned spin components on the templates make them difficult to measure individually~\cite{GWTC3, GWTCPOP}.

Current observations of BBH mergers suggest that half of the individual spin magnitudes lie below $\chi_i \approx 0.26$ and are typically misaligned with the orbital angular momentum $\Vec{\textbf{L}}$ \cite{GWTCPOP}. Misaligned spins cause $\Vec{\textbf{L}}$ to precess and nutate around the total angular momentum $\Vec{\textbf{J}}$ whose direction remains approximately constant during the inspiral part of the CBC~\cite{Apostolatos,Zhao}. This spin and orbital precession play a key role in BBH dynamics~\cite{Apostolatos} by imprinting modulations to the amplitude and the phase of the GW signal~\cite{Cutler}.

These modulations enter the GW templates at the $1.5$PN and $2$PN order as spin-orbit and spin-spin coupling effects respectively~\cite{RevModPhys.52.299,Cutler,Apostolatos,Kidder}. State-of-the-art waveform model families used in GW data analysis, such as IMRPhenomXPHM~\cite{IMRPHENOM}, SEOBNRv4PHM~\cite{SEOB}, and NRSur7dq4~\cite{NSUR}, model these effects accurately. However, it is fairly challenging to analytically quantify how the two spin vectors generically affect the GW template. 
As current detectors cannot unambiguously constrain both spins, see e.g.,~\cite{Hoy}, the two spins are typically represented by combinations of the BBH masses and spins that approximate the effects. Two main such parameters are the effective inspiral spin $\chi_{\rm eff}$~\cite{Damour2001} which is conserved through the PN inspiral~\cite{Racine2008} of the coalescence and influences its length, and the effective precession spin $\chi_{\rm p}$~\cite{chip1, chip2, Gerosa2021} which approximates the in-plane spin component responsible for spin precession. The spin parameter $\chi_{\rm p}$ as used by the LIGO collaboration is shown to be inconsistent because it accounts for some precession effects while averaging out others \cite{Gerosa2021}.  The $\chi_{\rm p}$ parameter evolves during the inspiral; fixing its value to that at a reference frequency can cause one to underestimate the evidence for precession \cite{Gerosa2021}.

In our previous analysis~\cite{Ourpaper}, we developed a different set of phenomenological parameters that encompass the essential features of spin precession and nutation. These five parameters provide a geometric framework to characterize the various configurations in which the evolution of $\Vec{\textbf{L}}$ about $\Vec{\textbf{J}}$ impact the BBH dynamics. In our current analysis, we take one step forward and introduce a new toy model for the inspiral portion of the gravitational waveform that includes the precession amplitude $\theta_{LJ}$ and precession frequency $\Omega_{LJ}$ as parameters.  We then show how these parameters imprint a distinct signature on the template's various amplitude and phase terms.  Finally, we analyze the distinguishability of precession in the current LVK observing runs by calculating mismatches for different source systems and exploring the parameter space to draw meaningful conclusions.

\citeauthor{Fairhurst2020rho_p}~\cite{Fairhurst2020rho_p} introduced the `precession signal-to-noise ratio', $\rho_{\rm p}$ as the parameter quantifying the measurability of precession in \cite{Fairhurst2020rho_p} using the two-harmonic approximation \cite{Fairhurst2020twoharmonics}. Yet the analysis presented in \cite{Fairhurst2020rho_p, Fairhurst2020twoharmonics} folded the two-dimensional precession parameter space presented here into their single parameter $b$ that describes the opening angle of the cone that orbital angular momentum moves around the total angular momentum.
This may lead to inconsistencies not unlike those associated with $\chi_{\rm p}$, potentially affecting their results.  A more recent paper \cite{Hoy2024Precession} uses harmonic decomposition to search for precession in LVK events and finds substantial evidence for precession only in the event GW$200129\_065458$.

This paper begins with a review of post-Newtonian inspiral waveform formalism in Section~\ref{sec: section 2 basic formalism}.  In Section~\ref{sec: RP model}, we introduce regular-precession parameters based on a taxonomy of spin precession \cite{Ourpaper} and then build a model for the waveforms. In Section~\ref{sec: mismatch and SNR calculations}, we employ match-filtering techniques to distinguish a regularly precessing waveform from a non-precessing one and set up a distinguishability criterion based on \citeauthor{Lindblom}~\cite{Lindblom}. We also discuss signal-to-noise ratios and the Lindblom inequality for different source orientations and sky locations for a single detector.  We then calculate the fraction of waveforms in which precession can be identified as a function of redshifts for populations characterized by either fixed precession parameters or a distribution of spin orientations.  Although the precession signal-to-noise parameter $\rho_{\rm p}$ can also be used to shed light on the distinguishability of precession for a particular signal, we stick to the Lindblom criterion for this analysis.  Finally, we conclude in Section~\ref{sec: conclusions} with some closing remarks on future directions for the model.  Appendices provide more context on the various reference frames we employ, the PN approximations adopted for the precession parameters, the significance of the Lindblom criterion, and information about the release of the data and supporting codes used in this paper.

Throughout this paper, we assume a flat Universe with present-day Hubble constant $H_0 = 67.4$~km/s and matter density $\Omega_m = 0.315$ to determine the luminosity distance as a function of redshift~\cite{Planck2018}.

\section{Basic formalisms and notations} \label{sec: section 2 basic formalism}

In this section, we briefly review the inspiral waveform in the Newtonian, quadrupole-moment approximation \cite{Apostolatos,PhysRevD.47.2198, Takahashi_2003} and our taxonomy of spin precession \cite{Ourpaper}. We will incorporate the latter into the former in the following section to construct a toy model of a precessing frequency-domain gravitational waveform.

\subsection{Gravitational waveform in the quadrupole-moment approximation} \label{subs: non-precessing template}

In the frequency domain, a GW signal observed by a ground-based detector can be expressed as
\cite{Apostolatos,Cutler,Takahashi_2003}

\begin{equation} \label{eq: hf}
\tilde{h}(f) = F_+ \tilde{h}_+(f) + F_\times \tilde{h}_{\times}(f)\,,
\end{equation}
where $F_+$ and $F_\times$ are the detector beam-pattern coefficients and $h_+$ and $ h_\times$ are the projections of the GW field onto the two principal polarization states.

For an L-shaped GW detector, these beam-pattern coefficients can be expressed as \cite{PhysRevD.17.379,Apostolatos}

\begin{subequations} \label{eq: BeamPatt}
\begin{align}
\label{eq:F+}
F_+(\theta_S,\Phi_S,\psi) &= \frac{1}{2} (1+\cos^2\theta_S) \cos 2\Phi_S \cos 2\psi \\
&\qquad - \cos\theta_S \sin 2\Phi_S \sin 2\psi\,, \notag \\
&= C \cos(2\psi + \alpha)\,, \notag \\
\label{eq:Fx}
F_\times(\theta_S,\Phi_S,\psi) &= \frac{1}{2} (1+\cos^2\theta_S) \cos 2\Phi_S \sin 2\psi \\
&\qquad +\cos\theta_S \sin 2\Phi_S \cos 2\psi\,, \notag \\
&= C \sin(2\psi + \alpha)\,, \notag
\end{align}
\end{subequations}
where the spherical angles $\theta_S, \Phi_S$ specify the sky location $\hat{\Vec{N}}$ of the source with respect to the detector frame $\{\hat{\Vec{X}}_D, \hat{\Vec{Y}}_D, \hat{\Vec{Z}}_D\}$ defined with respect to the arms of the L-shaped detector.  This frame, along with the sky and source frames we will use later, is described in greater detail in Appendix~\ref{Appendix: 3frames}.

The polarization angle $\psi$ between the principal $+$ direction and the direction of constant azimuth is given by \cite{Apostolatos}
\begin{equation} \label{eq: polarization angle psi}
\psi =  \tan^{-1}\left[ \frac{\hat{\Vec{L}} \cdot \hat{\Vec{Z}}_D -(\hat{\Vec{L}} \cdot \hat{\Vec{N}})(\hat{\Vec{Z}}_D \cdot \hat{\Vec{N}})}{\hat{\Vec{N}} \cdot (\hat{\Vec{L}} \times \hat{\Vec{Z}}_D)} \right] \,,
\end{equation}
where $\hat{\Vec{L}}$ is the unit vector in the direction of the BBH orbital angular momentum.  The beam-pattern amplitude and phase are
\begin{subequations} \label{eq: BeamPatt_amp_phase}
\begin{align}
C &\equiv \left[ \frac{1}{4} (1+\cos^2\theta_S)^2 \cos^2 2\Phi_S + \cos^2\theta_S \sin^2 2\Phi_S \right]^{1/2}, \label{eq: BeamPattamp} \\
\alpha &\equiv \tan^{-1}\left( \frac{2\cos\theta_S \tan 2\Phi_S}{1+\cos^2\theta_S} \right) \,. \label{eq: BeamPattphase} \,
\end{align}
\end{subequations}
The beam-pattern amplitude has a range $0 \leq C \leq 1$ and saturates the upper (lower) inequality for $\hat{\Vec{N}} = \pm\hat{\Vec{Z}}_D$ [$\hat{\Vec{N}} = 2^{-1/2}(\pm\hat{\Vec{X}}_D \pm\hat{\Vec{Y}}_D)$].

The frequency-domain GW polarizations are \cite{Cutler,Takahashi_2003}
\begin{subequations} \label{eq: GWstrains}
\begin{align}
\tilde{h}_+(f) &= Af^{-7/6} [ 1 + (\hat{\Vec{L}} \cdot \hat{\Vec{N}})^2 ] e^{i\Psi}\,, \label{eq:hplus frequency} \\
\tilde{h}_\times(f) &= -2iAf^{-7/6}(\hat{\Vec{L}} \cdot \hat{\Vec{N}})e^{\mathit{i}\Psi}\,, \label{eq:hcross frequency} \,
\end{align}
\end{subequations}
where

\begin{equation} \label{eq:takahasi amplitude}
A = \sqrt{\frac{5}{96}} \frac{1}{D_L}\frac{\mathcal{M}^{5/6}}{ \pi^{2/3}}\,,
\end{equation}
$D_{L}$ is the luminosity distance between the detector and the source, and $\mathcal{M}$ is the chirp mass.  For a binary with total mass $M = m_1 + m_2$ and mass ratio $q = m_2/m_1 \leq 1$, the symmetric mass ratio is $\eta \equiv q/(1+q)^2$ and the chirp mass is $\mathcal{M} \equiv \eta^{3/5}M$.

The spin-independent GW phase at 2PN order is \cite{PhysRevD.52.848}
\begin{align}
\label{eq:phase 2PN}
\Psi(f) &= 2 \pi f t_c - \phi_c - \frac{\pi}{4} + \frac{3}{128} \eta^{-1} x^{-5/2} \notag \\
&\qquad \times \left[ 1 + \frac{20}{9} \left(\frac{743}{336} + \frac{11}{4}\eta \right) x - 16\pi x^{3/2} + 10\Gamma x^2 \right]\,
\end{align}
where $x \equiv (\pi Mf)^{2/3}$ is the traditional PN parameter, $t_c$ and $\phi_c$ are the time and GW phase at binary coalescence, and
\begin{equation}
\Gamma \equiv \frac{3058673}{1016064}+\frac{5429}{1008}\eta+\frac{617}{114}\eta^2 \,.
\end{equation}

Inserting Eqs.~(\ref{eq: BeamPatt}) and (\ref{eq: GWstrains}) into (\ref{eq: hf}) yields
\begin{equation} \label{eq:hf A and phase}
\tilde{h}(f) = B e^{\mathit{i}[\Psi(f) - \phi_p - 2\delta\Phi]} \,,
\end{equation}

where the GW amplitude $B$ is given by
\begin{align} 
B &= Af^{-7/6} \{ 4(\hat{\Vec{L}} \cdot \hat{\Vec{N}})^2 F^2_\times + [1+(\hat{\Vec{L}}\cdot\hat{\Vec{N}})^2 ]^2 F^2_+ \}^{1/2}\,, \label{eq:hf amplitude} \notag \\
&= ACf^{-7/6} \{ 4(\hat{\Vec{L}} \cdot \hat{\Vec{N}})^2 \sin^2(2\psi + \alpha) \\
&\qquad + [1+(\hat{\Vec{L}}\cdot\hat{\Vec{N}})^2 ]^2 \cos^2(2\psi + \alpha) \}^{1/2} \,, \notag
\end{align}
the polarization phase $\phi_p$ is

\begin{align} 
\phi_p &= \tan^{-1}\left[ \frac{2(\hat{\Vec{L}} \cdot \hat{\Vec{N}}) F_\times}{[1 +(\hat{\Vec{L}} \cdot \hat{\Vec{N}})^2] F_+} \right] \,, \label{eq: phip} \\
&= \tan^{-1}\left[ \frac{2(\hat{\Vec{L}} \cdot \hat{\Vec{N}}) \tan(2\psi + \alpha)}{1 +(\hat{\Vec{L}} \cdot \hat{\Vec{N}})^2} \right]\,, \notag
\end{align}
and $\delta\Phi$ is an additional contribution to the GW phase for precessing BBH systems given by integrating \cite{Apostolatos} 
\begin{equation} \label{eq: delta correction}
\frac{d\delta\Phi}{df} = \left[ \frac{\hat{\Vec{L}} \cdot \hat{\Vec{N}}}{1 -(\hat{\Vec{L}} \cdot \hat{\Vec{N}})^{2}} \right] (\hat{\Vec{L}} \times \hat{\Vec{N}}) \cdot \frac{\hat{d\Vec{L}}}{df}\,.
\end{equation}
As we are only interested in the inspiral portion of the waveform during which precession can occur, we cut off our waveforms at a GW frequency
\begin{equation} \label{eq: fcut}
f_{\rm cut} = \frac{1}{6^{3/2}\pi M} = 4.3 \times 10^3\,{\rm Hz} \left( \frac{M}{M_\odot} \right)^{-1}
\end{equation}
equivalent to the quadrupole frequency at the innermost stable circular orbit of a BH of mass $M$ \cite{1992ApJ...400..175B,Cutler}. 

\subsection{Taxonomy of BBH precession and nutation} \label{subs: taxonomy}

\begin{figure}[t!]\label{fig: orbital frame}
    \centering
\includegraphics[scale = 0.5]{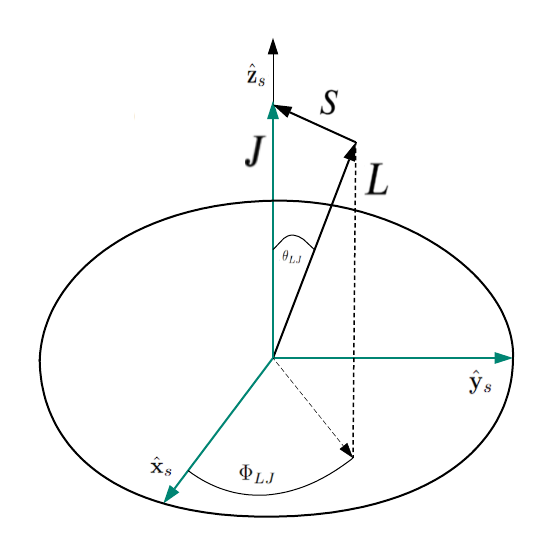}
\includegraphics[scale = 0.5]{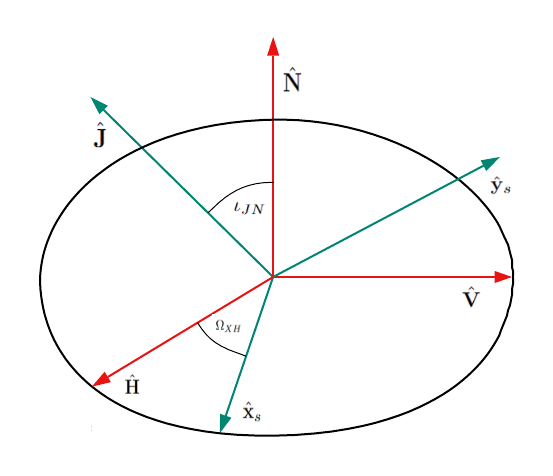}
    \caption{
    Schematic depictions of the sky and source coordinate frames.
    The sky frame, shown in red in the bottom panel, is defined such that the unit vector $\hat{\Vec{N}}$ pointing from the detector to the sky location of the GW source is along the $z$ axis, the unit vector $\hat{\Vec{V}}$ points towards the zenith at the detector location is along the $y$ axis, and the unit vector $\hat{\Vec{H}} = \hat{\Vec{V}} \times \hat{\Vec{N}}$ completes the orthonomal triad.
    The source frame is defined such that the total angular momentum $\hat{\Vec{J}}$ is along the $Z_S$ axis, $\hat{\Vec{N}} \times \hat{\Vec{J}}$ is along the $X_S$ axis, and $\hat{\Vec{Y}}_S = \hat{\Vec{Z}}_S \times \hat{\Vec{X}}_S$ completes the orthonormal triad.  
}
\end{figure}

The gravitational waveform described in the previous subsection can be determined once the frequency-dependent direction of the orbital angular momentum $\hat{\Vec{L}}(f)$ is specified. This problem was considered in \citeauthor{Ourpaper}~\cite{Ourpaper} which created a taxonomy of spin precession. Following \cite{Apostolatos}, this work defined precession as \emph{simple} when the direction of the total angular momentum $\hat{\Vec{J}}$ is approximately constant during the inspiral.  This is an excellent approximation except in the special cases of transitional precession \cite{Apostolatos} or nutational resonances \cite{Zhao}. In the generic case of simple precession, misaligned BBH spins make the orbital angular momentum $\Vec{L}$ precess about the total angular momentum $\Vec{J}$ with polar angle $\theta_{LJ}$ and azimuthal angle $\Phi_{LJ}$ that both vary on the precession timescale. The oscillation of $\theta_{LJ}$ on the precession timescale is called \emph{nutation}. In the special case of regular precession, nutation vanishes and both $\theta_{LJ}$ and the precession frequency $\Omega_{LJ} \equiv d\Phi_{LJ}/dt$ vary on the longer radiation-reaction timescale.

Nine general coordinates would be needed to specify $\Vec{L}$ and the individual spins $\Vec{S}_1$ and $\Vec{S}_2$ of a precessing BBH system.  However, for the case of simple precession, \citeauthor{PhysRevLett.114.081103}~\cite{PhysRevLett.114.081103} recognized that, on the precession timescale, the constancy of $\Vec{J}$, $L$, $S_1$, $S_2$, and the projected effective spin $\chi_{\rm eff}$ \cite{Damour2001, Racine2008} implies that such systems only possess two degrees of freedom.  These degrees of freedom can be described by one intrinsic parameter, the total spin magnitude $S = |\Vec{S}_1 + \Vec{S}_2|$, and one extrinsic parameter, a global rotation of the BBH system by an angle $\Phi_{LJ}$ about $\Vec{J}$.  As indicated in the top panel of Fig.~\ref{fig: orbital frame}, the oscillation of $S$ between its extrema $S_-$ and $S_+$ on the precession timescale corresponds to nutation, the oscillation of 
\begin{equation} \label{eq:ThetaL}
\theta_{LJ}(S) = \cos^{-1} \left( \frac{J^2 + L^2 - S^2}{2JL} \right)
\end{equation}
between its extrema $\theta_{LJ\pm} = \theta_{LJ}(S_\pm)$.  The duration of this oscillation defines the nutation period
\begin{equation} \label{eq:precessional period}
    \tau = 2\int_{S_{-}}^{S_{+}}\frac{dS}{|dS/dt|}\,.
\end{equation}
As $S$ describes the only intrinsic degree of freedom of the system, one can define the \emph{precession average} of any quantity $X$ varying on the precession timescale as
\begin{equation} \label{eq:precession average}
\langle X \rangle \equiv \frac{2}{\tau} \int_{S_-}^{S_+} \frac{X(S)}{|dS/dt|} dS\,.
\end{equation}
In addition to the polar angle $\theta_{LJ}(S)$, such quantities include the precession frequency $\Omega_{LJ}(S) = d\Phi_{LJ}/dt$.  Precessional dynamics in this formalism are described in much greater detail in \citeauthor{PhysRevD.92.064016}~\cite{PhysRevD.92.064016} which also provides expressions for $S_\pm$, $dS/dt$, and $\Omega_{LJ}(S)$.

\citeauthor{Ourpaper}~\cite{Ourpaper} used this formalism to describe the evolution of $\hat{\Vec{L}}$ in terms of five new precession parameters in the generic case of simple precession:
(1) the precession amplitude $\langle\theta_{LJ}\rangle$, (2) the precession frequency $\langle\Omega_{LJ}\rangle$, (3) the nutation amplitude $\Delta\theta_{LJ} = (\theta_{LJ+} - \theta_{LJ-})/2$, (4) the nutation frequency $\omega = 2\pi/\tau$, and (5) the variation of the precession frequency $\Delta\Omega_{LJ} = [\Omega_{LJ}(S_+) - \Omega_{LJ}(S_-)]/2$.

These parameters have several advantages over other approaches to characterizing generic simple precession.  By describing the evolution of $\hat{\Vec{L}}$, they have a more direct connection to the inspiral waveform (\ref{eq:hf A and phase}) than the individual spins $\Vec{S}_i$.  They also vary on the longer radiation-reaction timescale, making parameter estimates less sensitive to the reference frequency at which they are specified.  They also provide a richer and more accurate description of generic simple precession than any single parameter like the effective precession spin $\chi_{\rm p}$ \cite{chip1, chip2, Gerosa2021}. Even the case of regular precession considered in this paper is two dimensional, being described by the precession amplitude $\langle\theta_{LJ}\rangle$ and frequency $\langle\Omega_{LJ}\rangle$. 

In the next section, we incorporate these parameters into Eq.~(\ref{eq:hf A and phase}) to create a toy model of a regularly precessing waveform.

\section{A new model of regularly precessing waveforms} \label{sec: RP model}

We are now ready to construct our regularly precessing waveform.  We need to specify the direction of the orbital angular momentum $\hat{\Vec{L}}$ as a function of GW frequency $f$ and our precession parameters $\langle\theta_{LJ}\rangle$ and $\langle\Omega_{LJ}\rangle$.  Once this is determined, the expression $\hat{\Vec{L}}(f)$ can be inserted into the polarization angle $\psi$ of Eq.~(\ref{eq: polarization angle psi}), the GW amplitude $B$ of Eq.~(\ref{eq:hf amplitude}), and the GW phase $\phi_p + 2\delta\Phi$ of Eqs.~(\ref{eq: phip}) and (\ref{eq: delta correction}) to obtain the waveform $\tilde{h}(f)$.  The direction of the orbital angular momentum is given by
\begin{equation} \label{eq: LhatS}
\hat{\Vec{L}} = \sin\theta_{LJ}[\cos\Phi_{LJ} \hat{\Vec{X}}_S + \sin\Phi_{LJ} \hat{\Vec{Y}}_S] + \cos\theta_{LJ} \hat{\Vec{Z}}_S\,,
\end{equation}
where the \emph{source frame} $\{\hat{\Vec{X}}_S, \hat{\Vec{Y}}_S, \hat{\Vec{Z}}_S\}$ is defined such that $\hat{\Vec{Z}}_S = \hat{\Vec{J}}$ and $\hat{\Vec{X}}_S$ is along the line of ascending node of the plane perpendicular to $\hat{\Vec{J}}$ with respect to the plane of the sky.  This frame is depicted in Fig.~\ref{fig: orbital frame} and described in greater detail in Appendix~\ref{Appendix: 3frames}.

The precession parameters $\langle\theta_{LJ}\rangle$ and $\langle\Omega_{LJ}\rangle$ defined in the previous section are constant on the precession timescale, but 
we must make some assumption about their dependence on the GW frequency $f$ as the BBH inspirals through the sensitivity band of the GW detector.  For simplicity, we assume that these parameters maintain their lowest PN order frequency dependence as given by Eqs.~(9) and (10) of \cite{Ourpaper} all the way to merger,
\begin{subequations} \label{E:PPfreqdep}
\begin{align} 
\langle \theta_{LJ} \rangle &= \frac{0.1\tilde{\theta}}{4 \eta}\left(\frac{f}{f_{\rm cut}}\right)^{1/3}\,, \label{eq: thetaLJ} \\
\langle\Omega_{LJ} \rangle &= 10^3\,{\rm Hz}~\tilde{\Omega} \left(\frac{f}{f_{\rm cut}}\right)^{5/3}\left(\frac{M}{M_{\odot}}\right)^{-1}. \label{eq: Omega_LJ}
\end{align}
\end{subequations}

These equations implicitly define the dimensionless precession amplitude $\tilde{\theta}$ and frequency $\tilde{\Omega}$ which are constants of order unity.  
Appendix~\ref{Appendix: PN validations} assesses the validity of using the lowest PN order frequency dependence in Eqs.~(\ref{E:PPfreqdep}) along with our choice of normalizing the precession amplitude to be inversely proportional to the symmetric mass ratio.
We directly equate $\theta_{LJ}$ in Eq.~(\ref{eq: LhatS}) with $\langle \theta_{LJ} \rangle$ in Eq.~(\ref{eq: thetaLJ}), while $\Phi_{LJ}$ is obtained by integrating $\langle \Omega_{LJ} \rangle$ in Eq.~(\ref{eq: Omega_LJ}),
\begin{equation} \label{eq: PhiLJ}
\Phi_{LJ} = \gamma_p + \int^f_{f_{\rm min}} \langle\Omega_{LJ}\rangle \left( \frac{df'}{dt} \right)^{-1} df'~,
\end{equation}
where $f_{\rm min} = 20$\,Hz is the floor of the LVK sensitivity band, the integration constant $\gamma_p$ is the value of $\Phi_{LJ}$ when the binary enters this sensitivity band, and
\begin{align}
\frac{df}{dt} = \frac{96}{5} \pi\eta f^2 x^{5/2} \left[ 1 - \left( \frac{743}{336} + \frac{11}{4}\eta \right)x + 4\pi x^{3/2} \right]
\end{align}
is the rate at which the GW frequency increases during the inspiral at 1.5 PN order \cite{Cutler}.

\begin{figure*}[t!] \label{fig: cosL}
    \centering
    \includegraphics[width = \textwidth]{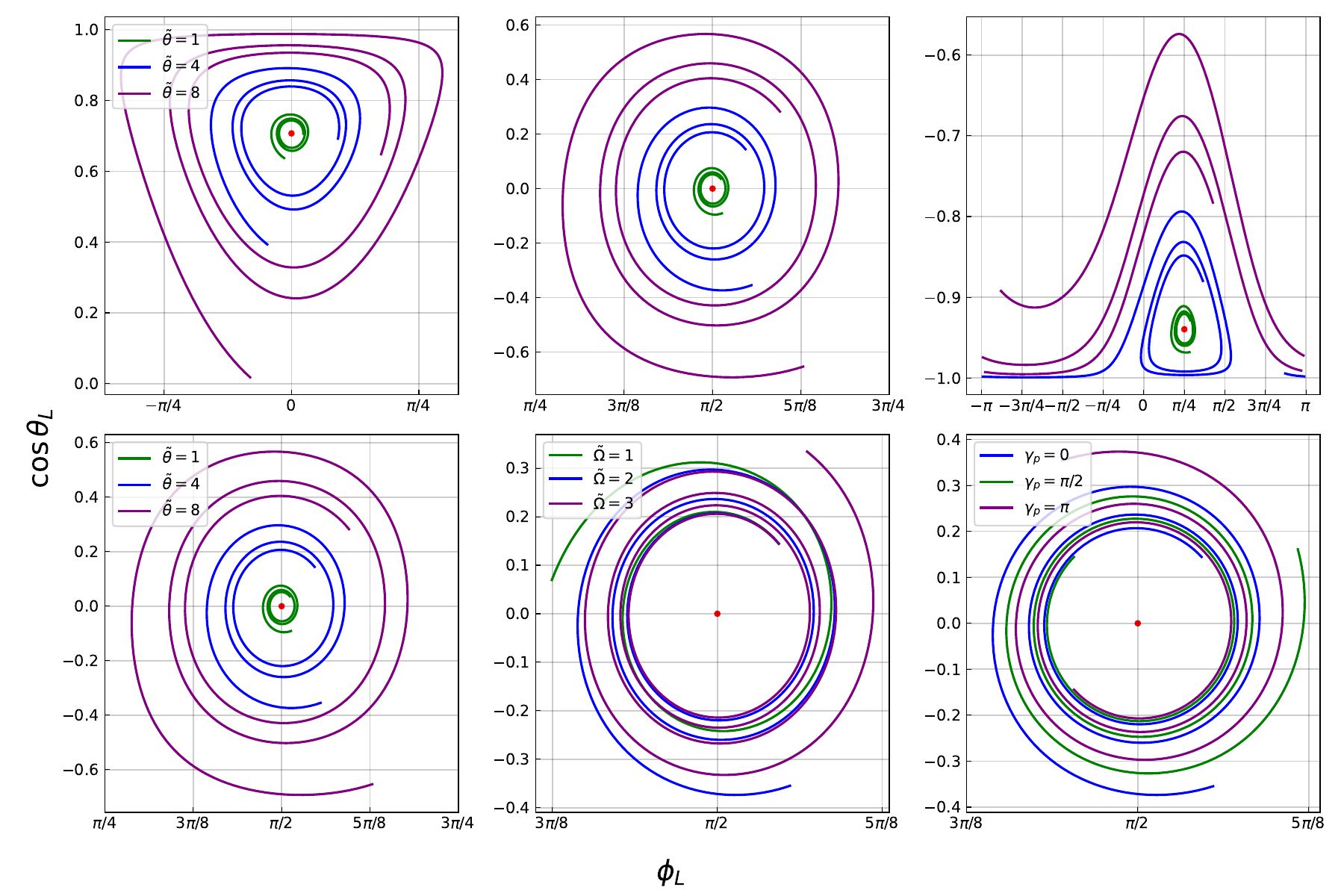}
    \caption{Evolution of the spherical coordinates $\theta_L$ and $\Phi_L$ that specify the direction of orbital angular momentum $\Vec{L}$ with respect to the detector frame.  The BBHs have equal masses with chirp mass $\mathcal{M} = 10 M_\odot$ and are at a cosmological redshift of $z = 0.3~(\simeq 1.5~\rm{Gpc~in~standard}~\Lambda{\rm CDM~cosmology})$.  The top row shows this precession for dimensionless precession frequency $\tilde{\Omega}$ = 2, azimuthal integration constant $\gamma_p$ = 0, and dimensionless precession amplitude $\tilde{\theta}= 1$ (green), 4 (blue), and 8 (purple).  The left, middle, and right panels of the top row correspond to different directions of the total angular momentum $\Vec{J}$ listed as Systems 1, 2, and 3 in Table~\ref{tab: source parameters}.  The bottom row shows the precession of $\hat{\Vec{L}}$ for System 2.  The left panel is a copy of the middle panel of the top row, the middle panel has $\tilde{\theta} = 4$, $\gamma_p = 0$, and $\tilde{\Omega}= 1$ (green), 2 (blue), and 3 (purple), and the bottom panel has $\tilde{\theta} = 4$, $\tilde{\Omega} = 2$, and $\gamma_p = 0$ (green), $\pi/2$ (blue), and $\pi$ (purple). The red dot in all panels denotes the total angular momentum direction.
    }
\end{figure*}

\begin{figure*}[t!]\label{fig: Amplitude various p}
    \centering
    \includegraphics[width = \textwidth]{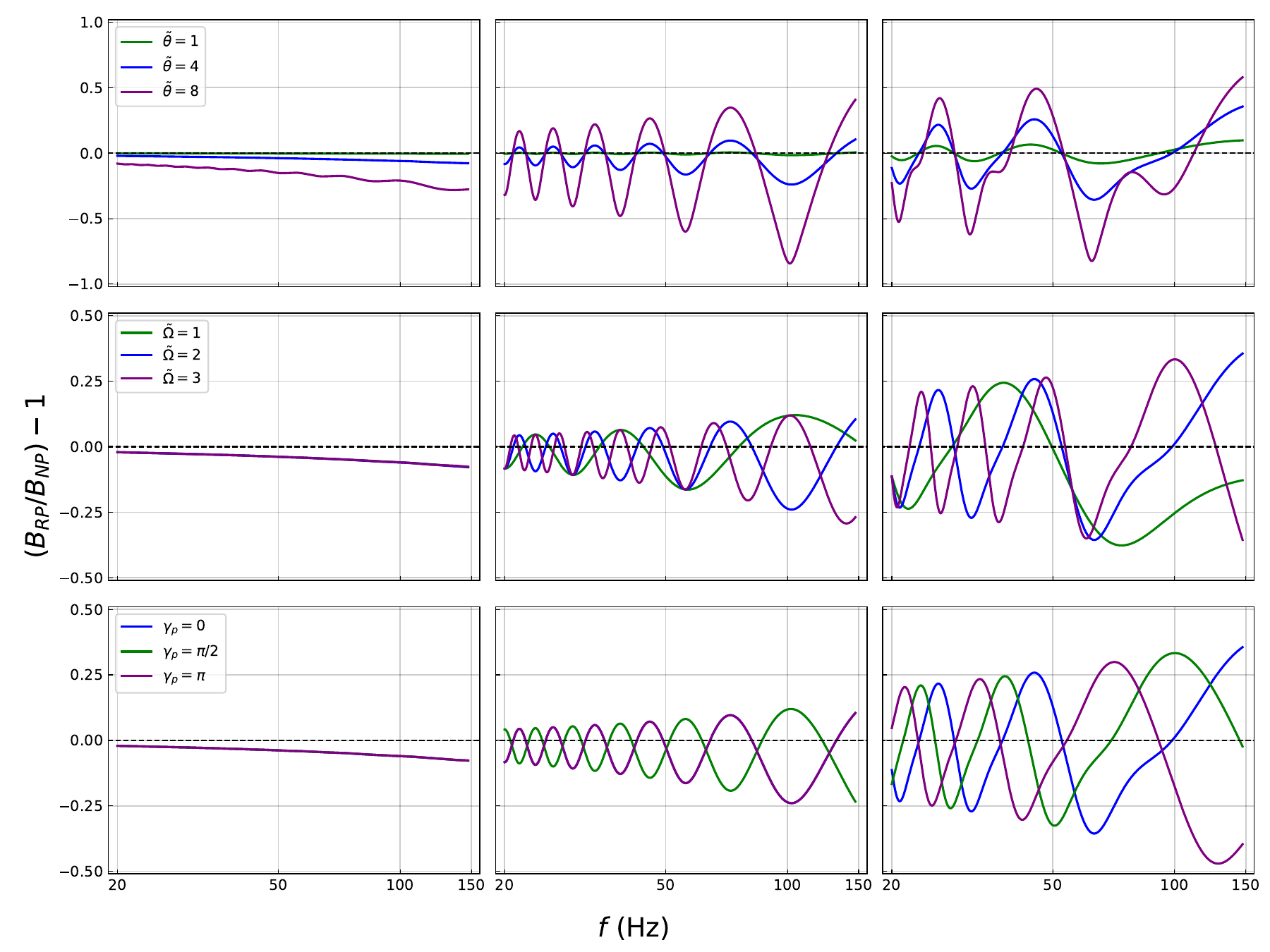}
    \caption{The ratio of the GW amplitude $B$, given by Eq.~(\ref{eq:hf amplitude}) in our regularly precessing waveform model, to the non-precessing waveform with the same BBH masses as in Fig.~\ref{fig: cosL} at redshift $z = 0.3$. The left, middle, and right panels correspond to directions of the total angular momentum $\hat{\Vec{J}}$ listed as Systems 1, 2, and 3 in Table~\ref{tab: source parameters}.  The top panels fix $\tilde{\Omega} = 2$, $\gamma_p = 0$ and vary $\tilde{\theta}$, the middle panels fix $\tilde{\theta} = 4$, $\gamma_p = 0$ and vary $\tilde{\Omega}$, and the bottom panels fix $\tilde{\theta} = 4$, $\tilde{\Omega} = 2$ and vary $\gamma_p$.}
\end{figure*}

\begin{figure*}[t!]\label{fig: phi_p plus 2 deltaphi p}
    \centering
    \includegraphics[width = \textwidth]{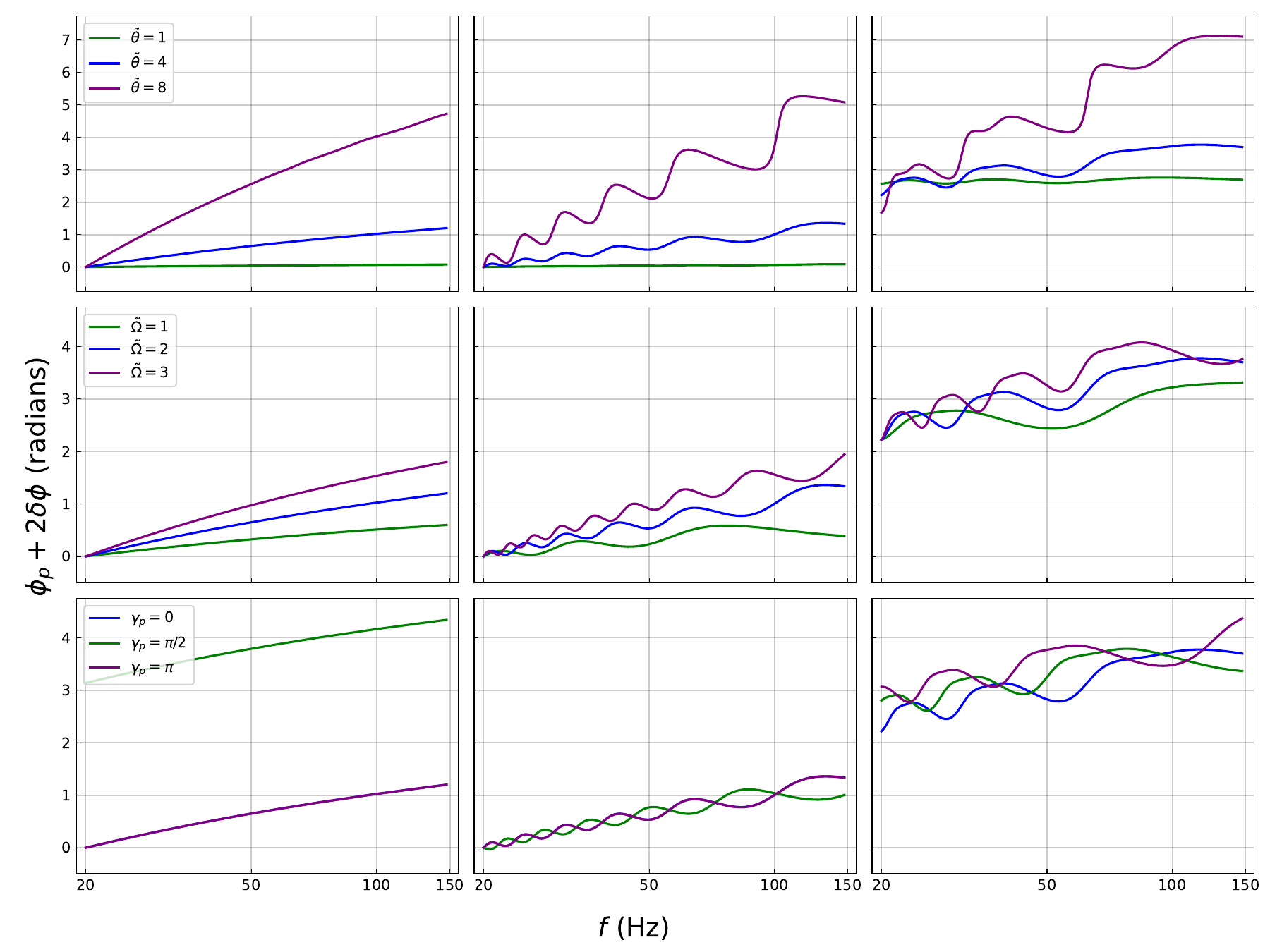}
    \caption{The evolution of the polarization phase and phase correction term $\phi_p + 2 \delta\Phi$ given by Eq.~(\ref{eq: phip}) and (\ref{eq: delta correction}) in our regularly precessing model (RP) versus the frequency with the same BBH masses as in Fig.~\ref{fig: cosL} at redshift $z = 0.3$. The left, middle, and right panels correspond to different orientations of the total angular momentum $\Vec{J}$ listed as Systems 1, 2, and 3 in Table~\ref{tab: source parameters}.  The top panels fix $\tilde{\Omega} = 2$, $\gamma_P = 0$ and vary $\tilde{\theta}$, the middle panels fix $\tilde{\theta} = 4$, $\gamma_P = 0$ and vary $\tilde{\Omega}$, and the bottom panels fix $\tilde{\theta} = 4$, $\tilde{\Omega} = 2$ and vary $\gamma_p$.}
\end{figure*}

We shall now explore this waveform model for three example BBH systems with different orientations for the total angular momentum $\Vec{J}$ specified by the values of $\theta_J$ and $\Phi_J$ listed in Table~\ref{tab: source parameters}. Systems 1 and 2 have face-on and edge-on inclinations, i.e. $\iota_{JN} = 0$ and $\pi/2$ respectively.  Note that we use the terms face-on and edge-on to refer to the fixed orientation between the total angular momentum and the line of sight, not that between the orbital angular momentum and the line of sight which varies as the BBHs precess and inspiral.
The parameters for System 3 are chosen to represent a generic case. These systems correspond to the three columns of panels in Figs.~\ref{fig: cosL} through \ref{fig: phi_p plus 2 deltaphi p} respectively showing the direction of the orbital angular momentum, the GW amplitude $B$, and the GW phase $\phi_p + 2\delta\Phi$.

\begin{table}[t!]
    \centering
    \begin{tabular}{ |c|c|c|c|  }
    \hline
     Parameter & System 1 & System 2 & System 3 \\
     \hline
     $\theta_J$ & $\pi/4$ & $\pi/2$ & $8\pi/9$ \\ \hline
     $\phi_J$ & 0 & $\pi/2$ & $\pi/4$ \\ \hline
     $\cos\iota_{JN}$ & 1 & 0 & -0.493 \\ \hline
     $\psi_0$ & --- & 0 & -1.29 \\ \hline
    \end{tabular}
    \caption{Spherical coordinates $\theta_J, \Phi_J$ specifying the direction of the total angular momentum $\Vec{J}$ in the detector frame, the inclination $\iota_{JN}$ between $\hat{\Vec{J}}$ and the sky location $\hat{\Vec{N}}$, and polarization angle $\psi_0$ for zero precession amplitude.  All three BBH systems have equal-mass BBHs with chirp mass $\mathcal{M} = 10M_{\odot}$ and sky location $\theta_S = \pi/4, \Phi_S = 0$.
    }
    \label{tab: source parameters}
 \end{table}

Fig.~\ref{fig: cosL} depicts the evolution of $\hat{\Vec{L}}$ according to Eqs.~(\ref{eq: LhatS}) - (\ref{eq: PhiLJ}) in the detector frame for equal-mass BBHs with chirp mass $\mathcal{M} = 10 M_\odot$.  

As the GW frequency $f$ increases from $f_{\rm min} = 20$\,Hz to $f_{\rm cut} = 146$\,Hz, $\hat{\Vec{L}}$ spirals in a cone about the direction of the total angular momentum $\hat{\Vec{J}}$ with increasing opening angle. The left, middle, and right panels in the top row of this figure show the three different directions $\hat{\Vec{J}}$ specified in the detector frame in Table~\ref{tab: source parameters}.  Each panel shows three choices of the dimensionless precession amplitude $\tilde{\theta}$ approximately equal to the $5^{\rm th}$, $50^{\rm th}$, and $95^{\rm th}$ percentiles of its distribution for equal-mass, maximally spinning, isotropically orientated BBHs which are respectively $\tilde{\theta} = 0.96$, 4.14, and 8.05, evaluated at the merger frequency $f_{\rm cut}$.  In the left, middle, and right panels of the bottom row of this figure, the direction $\hat{\Vec{J}}$ remains fixed to that of System 2, but the dimensionless precession amplitude $\tilde{\theta}$, dimensionless precession frequency $\tilde{\Omega}$, and azimuthal integration constant $\gamma_p$ are successively varied.  The three choices of $\tilde{\Omega}$ are approximately equal to the $5^{\rm th}$, $50^{\rm th}$, and $95^{\rm th}$ percentiles of its distribution for equal-mass, maximally spinning BBHs which are respectively $\tilde{\Omega} = 1.42$, 2.24, and 2.57, again evaluated at the merger frequency.
The effects of these changes on the precession of $\hat{\Vec{L}}$ are readily apparent; $\tilde{\theta}$ is proportional to the opening angle of the cone, $\tilde{\Omega}$ is proportional to the total angle precessed about the cone during the inspiral, and $\gamma_p$ corresponds to a global rotation of the system about $\hat{\Vec{J}}$.

Fig.~\ref{fig: Amplitude various p} shows how the the GW amplitude $B$ given by Eq.~(\ref{eq:hf amplitude}) depends on the dimensionless precession amplitude $\tilde{\theta}$, dimensionless precession frequency $\tilde{\Omega}$, and azimuthal integration constant $\gamma_p$ for equal-mass BBHs with chirp mass $\mathcal{M} = 10M_{\odot}$ and sky location $\theta_S = \pi/4, \Phi_S = 0$. This sky location implies a beam-pattern amplitude $C = 3/4$ and phase $\alpha = 0$ according to Eq.~(\ref{eq: BeamPatt_amp_phase}).
The left, middle, and right columns correspond to the three different choices of the direction of the total angular momentum $\Vec{J}$ listed as Systems 1, 2, and 3 in Table~\ref{tab: source parameters}. We plot the ratio of the regularly precessing (RP) to the non-precessing (NP) amplitude to focus on the effects of our precession parameters on the amplitude.

We see from the top panels of Fig.~\ref{fig: Amplitude various p} that as we increase the dimensionless precession amplitude $\tilde{\theta}$, the amplitude of the GW modulations increases as well. The middle row of Fig.~\ref{fig: Amplitude various p} shows that increasing $\tilde{\Omega}$ increases the number of precession cycles while leaving the GW amplitude unchanged. The integration constant $\gamma_p$ defined in Eq.~(\ref{eq: PhiLJ}) is the angle between the $\Vec{\hat{X}}_S$ axis in the source frame and the projection of $\Vec{L}$ onto the plane perpendicular to the total angular momentum $\Vec{J}$ when the BBH enters the sensitivity band of the detector ($f = f_{\rm min}$).  It is the precessional analogue to the orbital phase at coalescence $\phi_c$ and can similarly be regarded as a nuisance parameter.  Changing $\gamma_p$ shifts the phase of the GW amplitude modulations as seen in the bottom row of Fig.~\ref{fig: Amplitude various p}.

For the face-on system (System 1, $\iota_{JN}=0$) shown in the left column of Fig.~\ref{fig: Amplitude various p}, the GW amplitude is mildly suppressed according to Eq.~(\ref{eq:hf amplitude}) since $\hat{\Vec{L}} \cdot \hat{\Vec{N}} = \cos\theta_{LJ} \leq 1$. For the edge-on system (System 2, $\iota_{JN} = \pi/2$) shown in the middle column, $\hat{\Vec{L}} \cdot \hat{\Vec{N}} = \sin\theta_{LJ}\sin\Phi_{LJ}$ implying that the GW amplitude oscillates regularly over a precession period.
System 3 shown in the right column is a random choice of orientation providing an example of the more complicated generic dependence of the GW amplitude on frequency when both terms in Eq.~(\ref{eq:LN}) for $\hat{\Vec{L}} \cdot \hat{\Vec{N}}$ are nonzero.

Precession also modulates the GW phase through the two terms $\phi_p$ and $\delta\Phi$ given by Eqs.~(\ref{eq: phip}) and (\ref{eq: delta correction}). Fig.~\ref{fig: phi_p plus 2 deltaphi p} shows how these terms evolve for different orientations of the total angular momentum $\vec{J}$.  As was the case for the GW amplitude, the behavior of $\hat{\Vec{L}} \cdot \hat{\Vec{N}}$ during the inspiral largely determines the evolution of the GW phase. We note that there is a crucial difference between the evolution of the two contributions to the GW phase. The $\phi_p$ term is purely oscillatory unless the line of sight $\hat{\Vec{N}}$ lies inside the precession cone ($|\cos\iota_{JN}| \geq |\cos\theta_{LJ}|$), while the $\delta\Phi$ term always includes a secular component.

For the face-on case shown in the left column of Fig.~\ref{fig: phi_p plus 2 deltaphi p}, the oscillatory contribution to the GW phase is absent even for the largest precession amplitudes and frequencies.  Unlike the GW amplitude which only experiences oscillatory growth on the radiation-reaction timescale, the GW phase increases on the precession timescale with increasing values of both the precession amplitude and frequency contributing to this secular growth.  The edge-on case shown in the middle column shows regular oscillations whose amplitude and frequency increase respectively with the precession amplitude $\tilde{\theta}$ and frequency $\tilde{\Omega}$, as was the case for the GW amplitude.  System 3 shown in the right column provides an example of the more complicated frequency dependence expected for a generically oriented binary.

\section{Detecting precession as a function of binary orientation and sky location}
\label{sec: mismatch and SNR calculations}

In this section, we will assess the distinguishability of our regularly precessing (RP) waveforms from non-precessing (NP) waveforms using statistical estimates associated with matched filtering. We begin with a brief review of these matched-filtering techniques.

The $\emph{inner product}$ between two waveforms ${h}(f)$, ${g}(f)$ expressed in the frequency domain can be defined by
\begin{equation}\label{eq:inner product}
    \langle h|g \rangle = 4\Re \int_{f_{\rm min}}^{f_{\rm cut}}\frac{h(f) g^*(f)}{S_n(f)} df\,,
\end{equation}
where $S_n(f)$ is the power spectral density (PSD) of the detector noise \cite{PhysRevD.46.5236,PhysRevD.46.1517}.

This inner product can be used to compute the signal-to-noise ratio (SNR) of a source waveform $h_s$ detected using a template waveform $h_t$,~\cite{Cutler}:
\begin{eqnarray}
\label{eq:SNR}
\rho = \langle h_s|h_t \rangle^{1/2}.
\end{eqnarray}
We also define the $\emph{overlap}$
\begin{eqnarray}
\label{eq:overlap}
\mathcal{O}(h_s, h_t) = \frac{\langle h_s | h_t\rangle}{\sqrt{\langle h_s|h_s\rangle \langle h_t|h_t\rangle}}\,,
\end{eqnarray}
between the two signals; this overlap is normalized such that $\mathcal{O} \leq 1$, where the inequality is saturated for the overlap of a waveform with itself ($h_s = h_t$) \cite{Cutler}.

The $\emph{match}$ is defined as the overlap maximized over the time and the phase of coalescence:
\begin{eqnarray}
\label{eq:match}
   {\rm M}(h_s, h_t) = \max\limits_{t_c, \phi_c} \mathcal{O}(h_s, h_t) \,,
\end{eqnarray}
while the $\emph{mismatch}$ $\epsilon$ is
\begin{eqnarray}
\label{eq:mismatch}
\epsilon(h_s, h_t) = 1 - {\rm M}\,.
\end{eqnarray}
We use the \texttt{pycbc.filter} package \cite{alex_nitz_2022_6324278} to evaluate the mismatch between waveforms. For detector noise, we use the PSD for aLIGO (Advanced LIGO) from \cite{Sathyaprakash2009lrr}.

\begin{figure}[t!]
    \centering
    \includegraphics[width = \columnwidth]{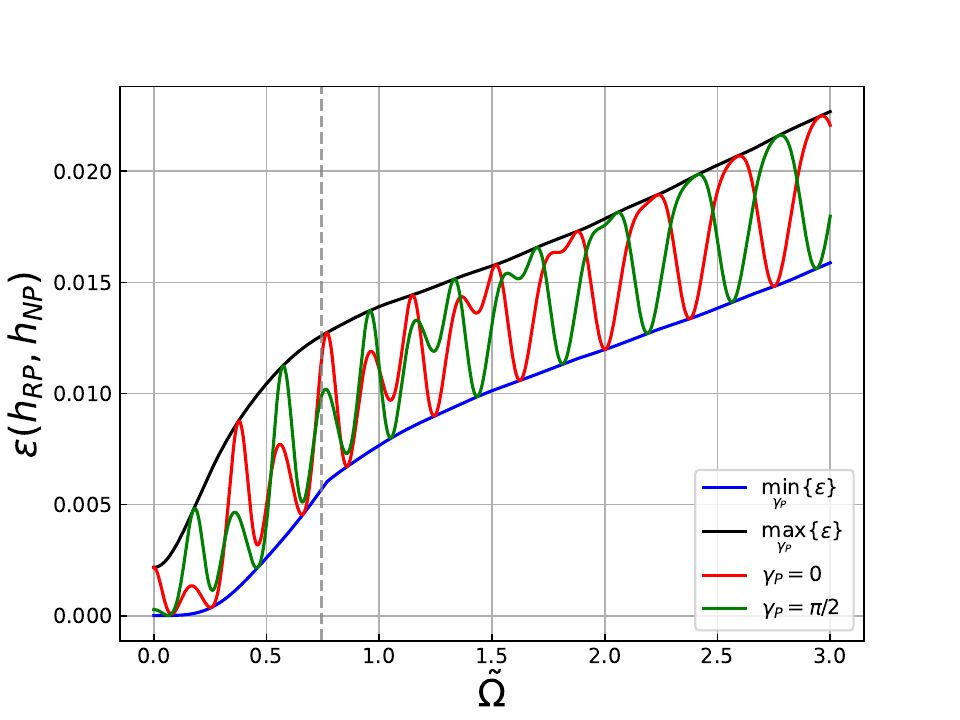}
    \caption{Mismatches between RP GW sources and NP templates as functions of the dimensionless precession frequency $\tilde{\Omega}$ for an equal-mass binary with chirp mass $\mathcal{M}=10M_{\odot}$.  The precession amplitude is fixed to be $\tilde{\theta} = 4$ and the sky location and binary orientation correspond to that of System 2 in Table~\ref{tab: source parameters}. We vary the integration constant $\gamma_p$ of Eq.~(\ref{eq: PhiLJ}) from $0$ to $2 \pi$ for each value of $\tilde{\Omega}$ and find the maximum (minimum) values of the mismatch shown by the black (blue) envelopes. The red and green curves are the mismatches for fixed $\gamma_p$ values of $0$ and $\pi/2$ respectively. The vertical dashed gray line denotes the value of $\tilde{\Omega}$ which corresponds to having one precession cycle as the BBH inspirals in the sensitivity band of the GW detector.}
    \label{fig: mismatch vs omega}
\end{figure}

Fig.~\ref{fig: mismatch vs omega} shows mismatches $\epsilon$ between an RP GW source and an NP template as functions of the dimensionless precession frequency $\tilde{\Omega}$.  For fixed values of the integration constant $\gamma_{p}$, the mismatch oscillates with $\tilde{\Omega}$ as the frequencies at which the crests and troughs of the GW amplitude and phase are located change with respect to the sensitivity bucket of the GW detector given by its PSD $S_n(f)$.  As $\gamma_{p}$ is a nuisance parameter with a flat distribution between 0 and $2\pi$, for the remainder of our analyses we make the conservative assumption to minimize the mismatch with respect to $\gamma_P$, defining
\begin{equation} \label{eq: mismatch gamma_p}
\epsilon_P = \min\limits_{\gamma_P} \epsilon (h_s, h_t)~.
\end{equation}
This assumption is shown by the lower blue envelope in Fig.~\ref{fig: mismatch vs omega}.

The vertical dashed gray line in Fig.~\ref{fig: mismatch vs omega} denotes the value of $\tilde{\Omega}$ at which one precession cycle occurs in the sensitivity band of the detector, i.e. as the GW frequency increases from $f_{\rm min}$ to $f_{\rm cut}$.  Below this value of the precession frequency, the minimum mismatch $\epsilon_p$ decreases rapidly since the fractional precession modulation can be moved to inconspicuous frequencies with respect to the sensitivity bucket through a judicious choice of $\gamma_p$.

\begin{figure*}[t!]\label{fig: opt mismatch for all systems}
    \centering
    \includegraphics[width = \textwidth]{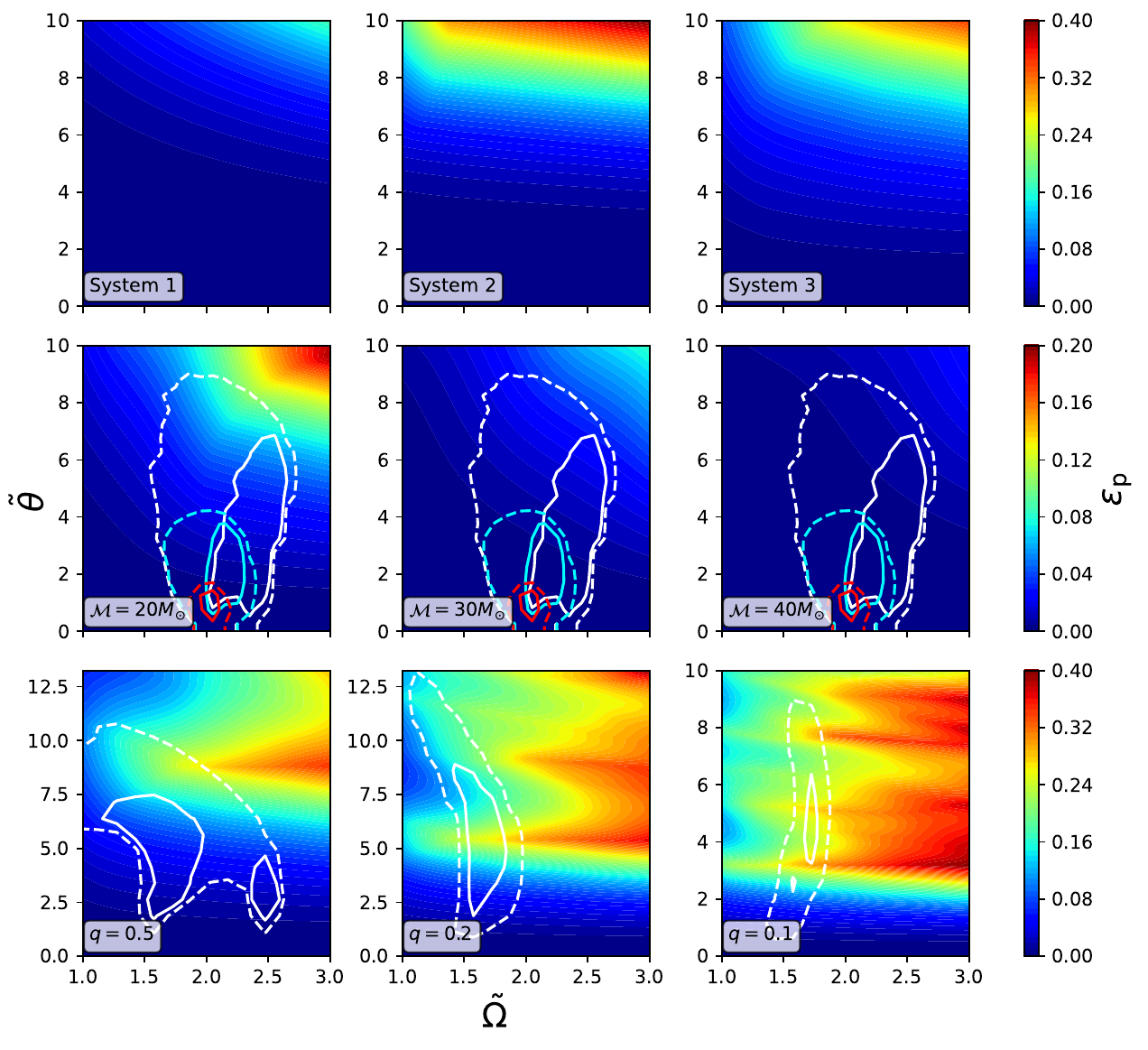}
    \caption{Mismatches between RP GW sources and NP templates minimized over $\gamma_p$ as a function of the precession frequency $\tilde{\Omega}$ and precession amplitude $\tilde{\theta}$. \textit{Top row}: The left, middle, and right panels correspond to Systems 1, 2, and 3 specified in Table~\ref{tab: source parameters}; the chirp mass is $\mathcal{M}=10M_{\odot}$. \textit{Middle row}: The left, middle, and right panels correspond to chirp masses of $\mathcal{M}=20M_{\odot}$, $\mathcal{M}=30M_{\odot}$, and $\mathcal{M}=40M_{\odot}$; the binary orientation is that of System 3. \textit{Bottom row}: The left, middle, and right panels correspond to mass ratios of $q=0.5$, $0.2$, and $0.1$; the chirp mass is $\mathcal{M}=10M_{\odot}$ and the binary orientation is that of System 3.  All BBHs are at redshift $z = 0.3$.  The solid (dashed) contours in the middle and bottom rows enclose 50\% (90\%) of the parameter distributions for isotropically oriented BBH spins with equal dimensionless magnitudes $\chi_i = 0.2$ (red), $0.5$ (cyan), and 1 (white).
    }
\end{figure*}

Fig.~\ref{fig: opt mismatch for all systems} shows contour plots of the minimized mismatches between RP sources and NP templates as functions of dimensionless precession amplitude $\tilde{\theta}$ and precession frequency $\tilde{\Omega}$. The top row corresponds to equal-mass binaries with a chirp mass of $\mathcal{M} = 10 M_\odot$.  System 1, the face-on orientation ($\hat{\Vec{J}} \parallel \hat{\Vec{N}}$), yields the lowest mismatches consistent with the conventional wisdom that it is hardest to detect precession for such orientations \cite{2014PhRvD..89l4025G}.
Conversely, it is easiest to identify precession in edge-on systems like System 2 shown in the middle panel.  This is to be expected from the large-amplitude oscillations of the GW amplitude and phase seen in the middle columns of Figs.~\ref{fig: Amplitude various p} and \ref{fig: phi_p plus 2 deltaphi p}. A more typical example of the detectability of precession may be provided by the generically oriented System 3 shown in the right panel.
For all three systems, the mismatches monotonically increase with increasing precession amplitude and frequency, peaking in the top right corner of each panel.

In the middle row of Fig.~\ref{fig: opt mismatch for all systems}, equal-mass systems in the orientation of System 3 increase in chirp mass from $\mathcal{M} = 20 M_\odot$ to $30 M_\odot$ and $40 M_\odot$ are shown in the left, middle, and right panels.  As the chirp mass increases, the BBH system coalesces at a lower frequency ($f_{\rm cut}$), reducing the time the signal spends in the sensitivity band of the detector. This implies fewer precession cycles observed for the same values of precession frequency $\tilde{\Omega}$, and, consequently, lower mismatches.  In all the panels of the middle row, contours of different colors indicate the distributions of the precession parameters for populations of equal-mass binaries with isotropically oriented spins with different spin magnitudes. The population with maximal spin magnitudes, shown in white, covers the largest area and most closely approaches the portion of parameter space with the highest mismatches in the top right corners.
As the spin magnitudes shrink to smaller values perhaps more typical of the systems observed during the third LIGO/Virgo observing run \cite{GWTCPOP} (cyan for $\chi = 0.5$ and red for $\chi = 0.2$), the distributions yield lower precession amplitudes and correspondingly lower mismatches.

In the bottom row of Fig.~\ref{fig: opt mismatch for all systems}, the mass ratio decreases from $q = 0.5$ to $0.2$ and $0.1$ in the left, middle, and right panels. As the mass ratio decreases, the number of precession cycles in band
\begin{equation} \label{E:Npre_est}
N_{\rm pre} \approx \left[ \langle \Omega_{LJ} \rangle \left( \frac{df}{dt} \right)^{-1} f \right]_{f_{\rm min}} \propto \eta^{-2/5} (\mathcal{M} f_{\rm min})^{-1}
\end{equation}
increases, leading to higher mismatches.  The left and middle panels in the bottom row have been extended to higher values of the precession amplitude $\tilde{\theta}$ than the top and middle rows to account for the larger misalignments between the orbital and total angular momenta that are possible at smaller mass ratios.  This has revealed one or more local maxima in the mismatch in all three panels.  Some of these maxima can be interpreted as systems in which the orbital angular momentum is aligned or anti-aligned with the line of sight at merger.  This occurs when $\langle \theta_{LJ} \rangle = \iota_{JN}$ or $\pi - \iota_{JN}$ at $f = f_{\rm cut}$.  For System 3, $\iota_{JN} = 2.09$ radians, implying that (anti-)alignment will occur for $\langle \theta_{LJ} \rangle = 2.09$ ($1.06$).  For $q=0.5$, $0.2$, and $0.1$, this corresponds to (anti-)alignment at $\tilde{\theta} = 18.5$ ($9.38$), $11.6$ ($5.87$), and $6.91$ ($3.49$) respectively.  The local maxima with the lowest values of $\tilde{\theta}$ do indeed appear to correspond to anti-aligned systems at $f = f_{\rm cut}$, since $\iota_{JN} > \pi/2$ for System 3.  Compared to the equal-mass systems seen in the top and middle rows, the unequal-mass systems in the bottom row have distributions of the precession parameters that extend more deeply into the high-mismatch regions, suggesting that identifying precession might be easier in unequal-mass systems.

\begin{figure*}[t!]\label{fig: SNR J and N}
    \centering
    \includegraphics[width = \textwidth]{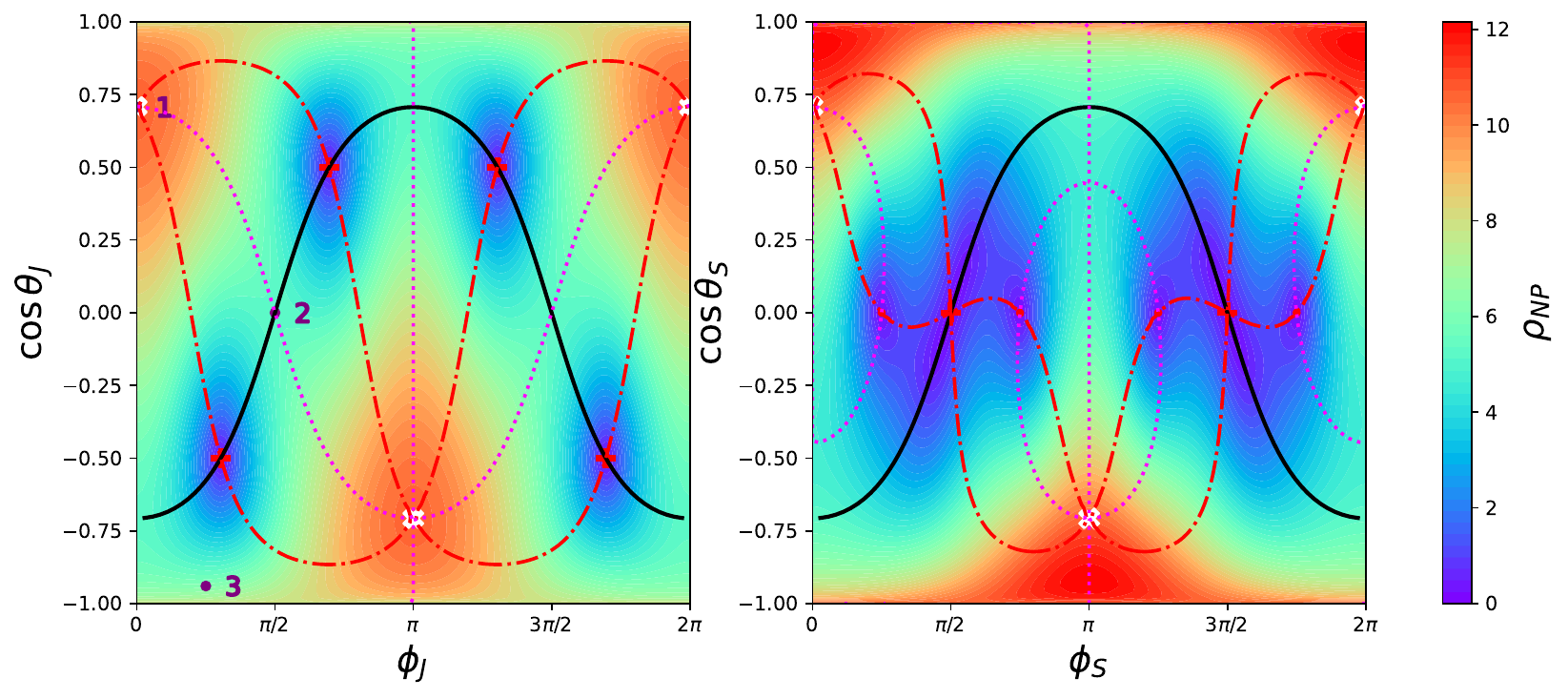}
    \caption{Signal-to-noise ratios $\rho$ for an NP source from a binary system with chirp mass $\mathcal{M}=10M_{\odot}$ at a redshift of $z = 0.3$. In the left panel, the line of sight $\hat{\Vec{N}}$ is given as $(\theta_S, \phi_S) = (\pi/4,0)$ with the orientation of the total angular momentum $\hat{\Vec{J}}$ for the binary varying as specified by
    the polar coordinates $(\theta_J, \phi_J)$. 
    For the right panel, the total angular momentum is fixed as $(\theta_J, \phi_J) = (\pi/4,0)$ and the system is located along the line of sight $\hat{\Vec{N}}$ specified by the
    polar coordinates $(\theta_S, \phi_S)$. 
    The solid black lines correspond to the family of edge-on orientations $(\hat{\Vec{J}} \cdot \hat{\Vec{N}} = 0)$ and the white crosses are face-on orientations $(\hat{\Vec{J}} \cdot \hat{\Vec{N}} = \pm 1)$. The dash-dotted red (dotted purple) curves are systems with $F_+ = 0$ ($F_\times = 0$). Intersections between the solid black and dash-dotted red lines are marked by red crosses and referred to as \emph{+~nulls}.  Intersections between the dash-dotted red and dotted purple lines are marked by red dots and referred to as \emph{true nulls}. The three systems listed in Table~\ref{tab: source parameters} are marked with numbered purple dots in the left panel.}
\end{figure*}

A source waveform $h_s$ can be distinguished from a template waveform $h_t$ for mismatches
\begin{align}
    \epsilon(h_{\rm s}, h_{\rm t}) \geq \frac{1}{2\rho^2}
    \label{eq: lindblom criterion}
\end{align}
where $\rho$ is SNR given in Eq.~(\ref{eq:SNR}) \cite{Lindblom}. This inequality is referred to as the Lindblom criterion or inequality. 

We first analyze SNRs for NP signals to understand the distinguishability of precession for different binary orientations and sky locations. Fig.~\ref{fig: SNR J and N} shows SNRs for different binary orientations at a fixed sky location ($\theta_S, \phi_S$)  = ($\pi/4, 0$) in the left panel and different sky locations and a fixed binary orientation ($\theta_J, \phi_J$) = ($\pi/4, 0$) in the right panel.

In both panels, the dash-dotted red (dotted purple) curves are systems with $F_+ = 0$ ($F_\times = 0$), while the solid black curves are edge-on systems only producing the $h_+$ polarization.  In the left panel, the SNR vanishes when $h_\times = F_+ = 0$ (intersection of the solid black and dash-dotted red curves).  In the right panel, the SNR vanishes for these systems as well as the true nulls located at $\theta_S = \pi, \Phi_s = (2n+1)\pi/4$ where $F_+ = F_\times = 0$ (intersection of the dash-dotted red and dotted purple curves).  In both panels, the SNR is maximized for face-on configurations $(\hat{\Vec{J}} \cdot \hat{\Vec{N}} = \pm 1)$ despite the coordinate singularities at these points erroneously suggesting that they are true nulls.

\begin{figure*}[t!]\label{fig: opt mismatch contours J varying}
    \centering
    \includegraphics[width = \textwidth]{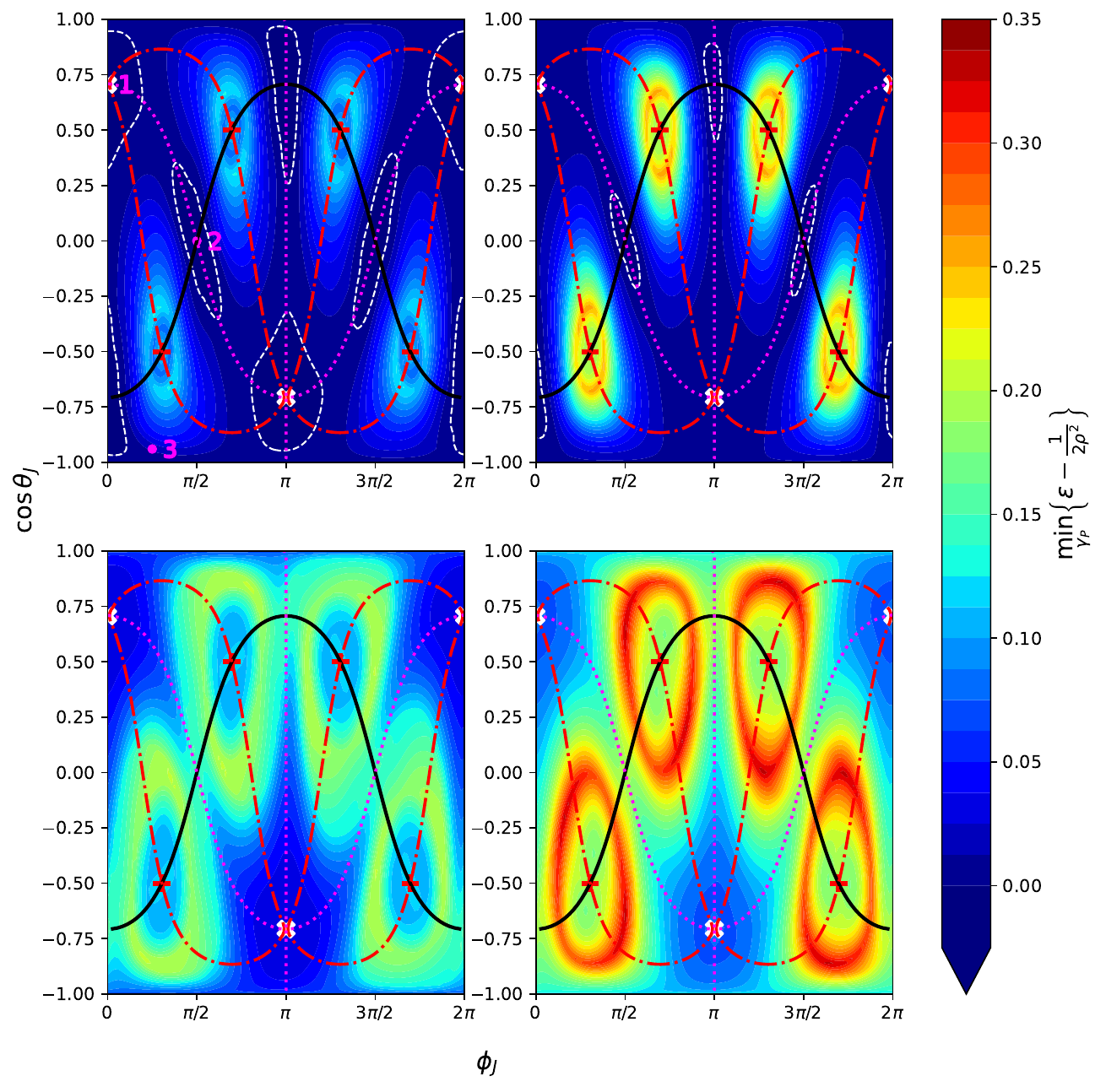}
    \caption{
    The quantity $\mathcal{I}_P$ of Eq.~(\ref{eq: lindblom criterion modified}) as a function of binary orientation ($\theta_J, \phi_J$) for merging BBHs with chirp mass of $\mathcal{M} = 10 M_\odot$ at redshift of $z = 0.3$ and fixed sky location ($\theta_S, \phi_S$) = ($\pi/4, 0$).  RP sources can be distinguished from NP templates whenever $\mathcal{I}_P > 0$, i.e. for areas outside the dashed white null contours.  As in Fig.~\ref{fig: SNR J and N}, the solid black, dot-dashed red, and dotted purple curves denote systems for which $\hat{\Vec{J}} \cdot \hat{\Vec{N}}$, $F_+$, and $F_\times$ vanish respectively, while the red (white) crosses denote +~nulls (face-on) systems.  The top left panel also shows the systems listed in Table~\ref{tab: source parameters}. The left (right) panels show sources with a precession frequency $\tilde{\Omega} = 2$ (3), and the top (bottom) panels show sources with a precession amplitude $\tilde{\theta} = 4$ (8).
    }
\end{figure*}

\begin{figure*}[t!]\label{fig: edge-on family contours}
    \centering
    \includegraphics[width = \textwidth]{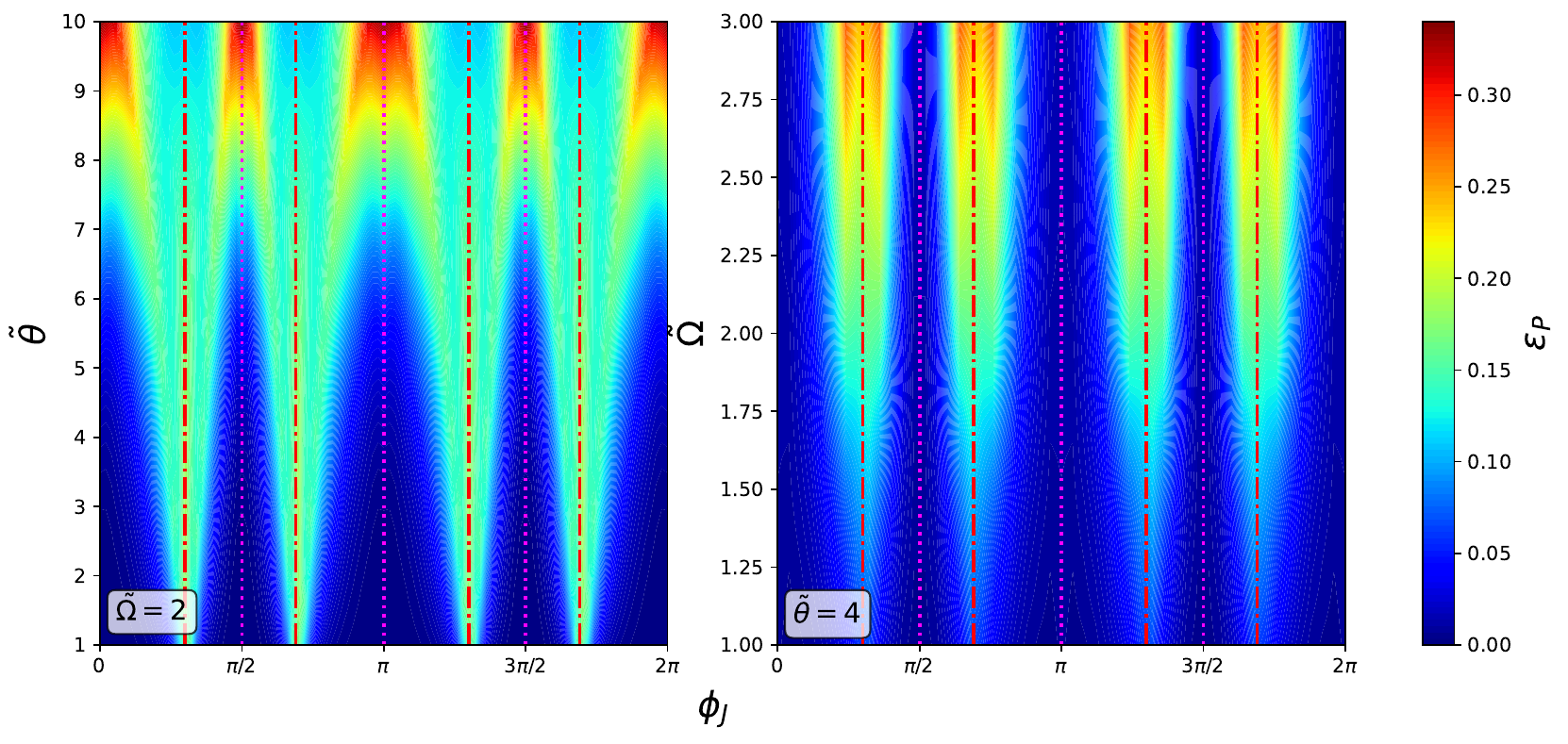}
    \caption{The mismatch $\epsilon_P$ between RP GW sources and NP templates minimized over precessional phase $\gamma_P$ for edge-on binaries $(\hat{\Vec{J}} \cdot \hat{\Vec{N}} = 0)$ with chirp mass of $\mathcal{M} = 10 M_\odot$ at sky location ($\theta_S, \phi_S$) = ($\pi/4, 0$) and redshift $z = 0.3$.  The horizontal axis shows the azimuthal coordinate $\phi_J$ specifying the direction of $\hat{\Vec{J}}$, while the vertical axis in the left (right) panel shows the precession amplitude $\tilde{\theta}$ (frequency $\tilde{\Omega}$).  In the left (right) panels, the precession frequency (amplitude) are fixed at $\tilde{\Omega} = 2$ ($\tilde{\theta} = 4$). The dot-dashed red (dotted purple) vertical lines indicate NP binaries for which $F_+ = 0$ ($F_\times = 0$) as in Figs.~\ref{fig: SNR J and N} and \ref{fig: opt mismatch contours J varying}.
    }
\end{figure*}

As $\gamma_{p}$ is a nuisance parameter with a flat distribution between 0 and $2\pi$, we can use the Lindblom criterion (\ref{eq: lindblom criterion}) to claim conservatively that a RP source with given precession amplitude and frequency can be distinguished from a NP template whenever the quantity

\begin{align}
    \mathcal{I}_P \equiv \min_{\gamma_P}\left\{\epsilon(h_{\rm s}, h_{\rm t}) - \frac{1}{2\rho^2} \right\} \geq 0~,
    \label{eq: lindblom criterion modified}
\end{align}

where we use the SNR $\rho$ for the RP source. A brief discussion on this choice can be found in Appendix~\ref{Appendix: Lindblom SNR discussion}. 

Fig.~\ref{fig: opt mismatch contours J varying} shows $\mathcal{I}_P$ for the GW signals from precessing binaries at a fixed sky location ($\theta_S = \pi/4, \Phi_S = 0$) with varying orientations.  All four panels of this figure look qualitatively similar; rings with high values of $\mathcal{I}_P$ surround the four red crosses denoting the +~nulls first seen in the left panel of Fig.~\ref{fig: SNR J and N}.  At these +~nulls, the signal vanishes for NP systems since $h_\times = F_+ = 0$.  However, the non-zero opening angle of the precession cone ($\tilde{\theta} \neq 0$) implies that these systems have non-vanishing signals for RP systems and correspondingly large mismatches with vanishing NP signals.  The angular radii of these rings correspond to the opening angle of the precession cone at the GW frequency $f_s$ of the bottom of the bucket of the LIGO sensitivity band.  Binaries with total angular momentum on these rings have orbital angular momenta that precess into the direction of the +~nulls at $f_s$, maximizing the precessional modulation of the GW signals and their mismatches with NP signals.  The angular radii of these rings are twice as large in the bottom panels compared to the top panels since $\tilde{\theta}$ changes from 4 to 8.  As the precession frequency $\tilde{\Omega}$ changes from 2 in the left panels to 3 in the right panels, the radii of the rings remain constant, but the mismatches increase because of the greater accumulation of the GW phase as seen in the middle row of Fig.~\ref{fig: phi_p plus 2 deltaphi p}.  Although the face-on configurations marked by the white crosses have the largest SNR and thus the least negative second terms in Eq.~(\ref{eq: lindblom criterion modified}) for $\mathcal{I}_P$, the lack of precession modulation as seen in the first columns of Figs.~\ref{fig: Amplitude various p} and \ref{fig: phi_p plus 2 deltaphi p} yield small mismatches making these the most difficult systems in which to identify precession.  It is also challenging to identify precession for binaries like System 2 located at the intersection of the solid black and dotted purple curves where $h_\times = F_\times = 0$.  Although precession generates a contribution to the $h_\times$ polarization for these signals at linear order in the precession amplitude $\tilde{\theta}$, this cannot be seen for these systems since $F_\times = 0$.

In Fig.~\ref{fig: edge-on family contours}, we examine the mismatches $\epsilon_P$ of edge-on binaries in greater detail.  In the left panel, we see that for small values of the precession amplitude $\tilde{\theta}$, the maxima (minima) of $\epsilon_P$ are indeed at orientations for NP binaries at which $F_+ = 0$ ($F_\times = 0$).  As $\tilde{\theta}$ increases, the values of $\Phi_J$ with large mismatches grow consistent with the wider rings in the bottom panels of Fig.~\ref{fig: opt mismatch contours J varying} compared to the top panels.  Large mismatches are only possible for binaries in the $F_\times = 0$ configurations at low GW frequencies $f$ if $\tilde{\theta}$ is large enough such that the opening angle of the precession cone can expand to the value ($\pi/4$) at which the orbital angular momentum can precess into a configuration for which $F_+ = 0$ by the end of the inspiral at $f = f_{\rm cut}$.  According to Eq.~(\ref{eq: thetaLJ}), this occurs at $0.1\tilde{\theta} = \pi/4$, i.e. $\tilde{\theta} \approx 7.85$ consistent with the peaks of the dark blue contours in the left panel of Fig.~\ref{fig: edge-on family contours}.  At still larger values of $\tilde{\theta}$, the locations of the maxima and minima of the mismatch $\epsilon_P$ are exchanged, with the maxima (minima) now located at values of $\Phi_J$ that correspond to $F_\times = 0$ ($F_+ = 0$) for NP binaries.  This occurs because these widely precessing binaries only spend the early portions of the inspiral near the corresponding NP configurations at which $F_\times = 0$ ($F_+ = 0$) and therefore the detector is not (is) sensitive to precession.  The GW detector is not very sensitive at these low frequencies.  By the time these widely precessing binaries inspiral to higher frequencies near the minima of the noise spectral density $S_n(f)$, their precession cones have already reached opening angles at which the angular momentum can pass through $F_+ = 0$ ($F_\times = 0$) for NP binaries leading to large (small) mismatches.

The right panel of Fig.~\ref{fig: edge-on family contours} is easier to interpret.  At a fixed value of $\Phi_J$, the mismatch monotonically increases with the precession frequency $\tilde{\Omega}$ as a greater difference in the GW phase accumulates during the inspiral as seen in the middle panel of Fig.~\ref{fig: phi_p plus 2 deltaphi p}.

\begin{figure*}[t!]\label{fig: opt mismatch contours N varying}
    \centering
    \includegraphics[width = \textwidth]{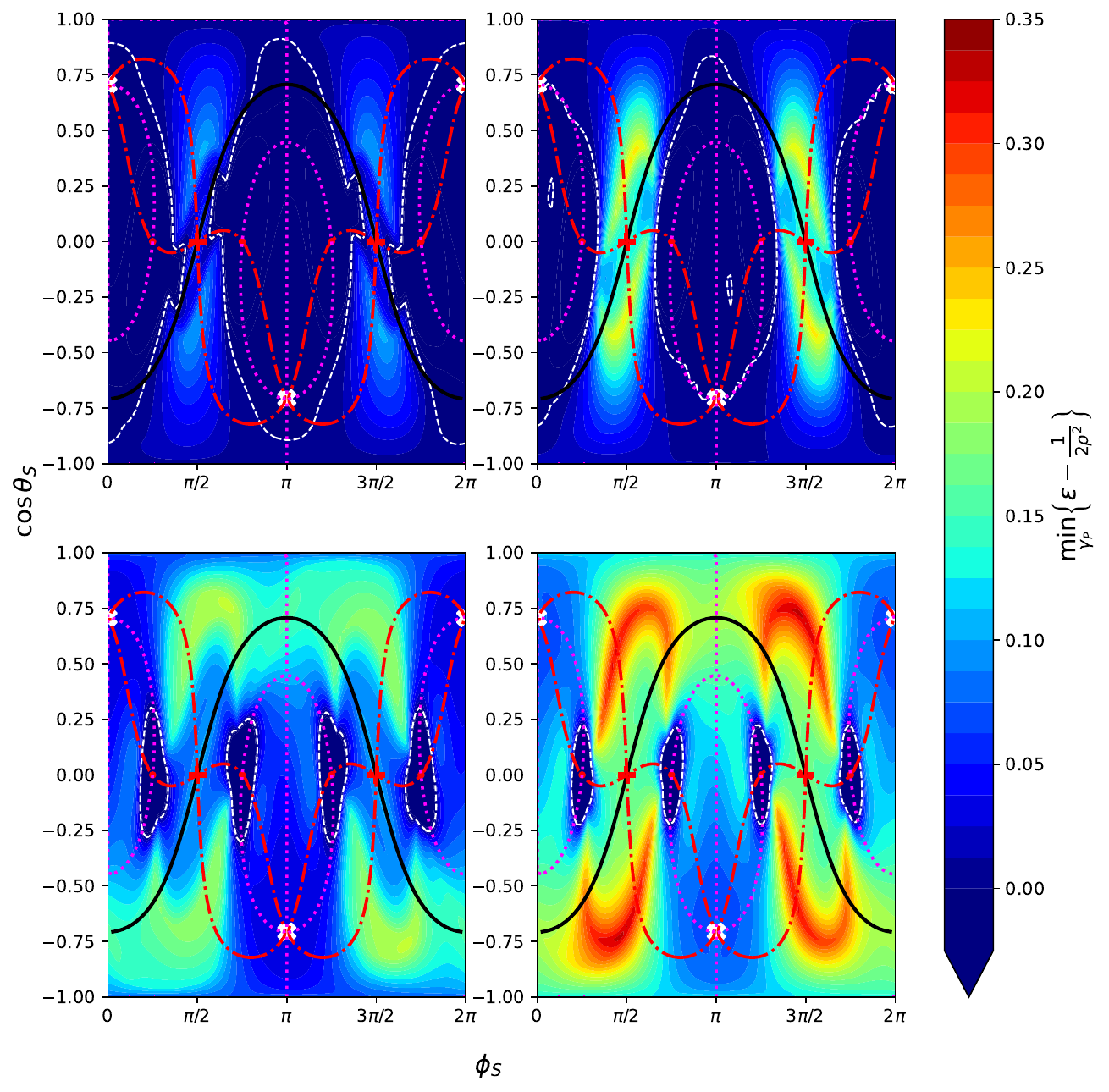}
    \caption{The quantity $\mathcal{I}_P$ of Eq.~(\ref{eq: lindblom criterion modified}) as a function of sky location ($\theta_S, \phi_S$) for merging BBHs with chirp mass of $\mathcal{M} = 10 M_\odot$ at redshift of $z = 0.3$ and fixed binary orientation ($\theta_J, \phi_J$) = ($\pi/4, 0$).  RP sources can be distinguished from NP templates whenever $\mathcal{I}_P > 0$, i.e. for areas outside the dashed white null contours.  As in Figs.~\ref{fig: SNR J and N} and \ref{fig: opt mismatch contours J varying}, the solid black, dot-dashed red, and dotted purple curves denote systems for which $\hat{\Vec{J}} \cdot \hat{\Vec{N}}$, $F_+$, and $F_\times$ vanish respectively, while the red (white) crosses denote +~nulls (face-on) systems.  The red dots denote true nulls at which $F_+ = F_\times = 0$.  The left (right) panels show sources with a precession frequency $\tilde{\Omega} = 2$ (3), and the top (bottom) panels show sources with a precession amplitude $\tilde{\theta} = 4$ (8).
    }
\end{figure*}

\begin{table*}
    \centering
    \begin{tabular}{|c|c|c|c|c|c|c|c|c|c|c|}
    \hline
    $\mathcal{M}$ & $q$ & $\tilde{\theta}$ & $\tilde{\Omega}$ & $z = 0.1$ & $0.2$ & $0.3$ & $0.4$ & $0.5$ & $0.75$ & $1$\\
    \hline
    \hline
    20 $M_{\odot}$& 1 & 4 & 2 & 97.3 & 86.8 & 64.5 & 37.0 & 15.4 & 0.3 & 0.0 \\
    \hline
    10 $M_{\odot}$& 1 & 4 & 2 & 97.9 & 89.6 & 72.7 & 50.6 & 29.5 & 4.0 & 0.0 \\
    \hline
    40 $M_{\odot}$& 1 & 4 & 2 & 85.4 & 39.2 & 8.2 & 0.3 & 0.0 & 0.0 & 0.0 \\
    \hline
    20 $M_{\odot}$& 0.5 & 4 & 2 & 97.4 & 85.7 & 61.8 & 33.0 & 12.8 & 0.1 & 0.0 \\
    \hline
    20 $M_{\odot}$& 0.1 & 4 & 2 & 91.4 & 56.5 & 16.8 & 1.7 & 0.0 & 0.0 & 0.0 \\
    \hline
    10 $M_{\odot}$& 0.1 & 4 & 2 & 97.2 & 83.9 & 54.3 & 25.9 & 8.7 & 0.0 & 0.0 \\
    \hline
    20 $M_{\odot}$& 1 & 1 & 2 & 57.8 & 8.1 & 0.2 & 0.1 & 0.0 & 0.0 & 0.0 \\
    \hline
    20 $M_{\odot}$& 1 & 8 & 2 & 99.5 & 96.9 & 89.9 & 76.3 & 56.6 & 11.8 & 0.6 \\
    \hline
    20 $M_{\odot}$& 1 & 4 & 1 & 93.2 & 63.9 & 30.0 & 10.1 & 2.2 & 0.0 & 0.0 \\
    \hline
    20 $M_{\odot}$& 1 & 4 & 3 & 98.6 & 91.5 & 75.8 & 51.7 & 29.2 & 1.8 & 0.0 \\
    \hline
    \end{tabular}
    \caption{Percentages of binaries with detectable precession from isotropically oriented populations with different chirp masses $\mathcal{M}$, mass ratios $q$, precession amplitudes $\tilde{\theta}$, and precession frequencies $\tilde{\Omega}$ indicated by the values in the first four columns.  The remaining columns show the percentages of systems with detectable precession for these populations at different redshifts $z$ with values indicated in the top row.  We use $10^4$ binaries for each row in this table leading to $\sim 1\%$-level errors due to Poisson noise.
    } \label{tab: MC on source and sky loc - fixed prec. param}
\end{table*}

We next examine the distinguishability of precession as a function of sky location in Fig.~\ref{fig: opt mismatch contours N varying}.  The interpretation of this figure is again guided by the solid black, dot-dashed red, and dotted purple curves that identify binary systems for which $\hat{\Vec{J}} \cdot \hat{\Vec{N}}$, $F_+$, and $F_\times$ vanish respectively.  As before, precession is most readily identified, i.e. $\mathcal{I}_P$ is largest, for RP systems in which the orbital angular momentum can precess into the +~nulls of the NP waveforms during the inspiral.  These systems consist of rings about the +~nulls with angular size proportional to the precession amplitude $\tilde{\theta}$, although there are only two of these rings in the plots of $\mathcal{I}_P$ as a function of sky location compared to the four rings in the plots of $\mathcal{I}_P$ as a function of binary orientation shown in Fig.~\ref{fig: opt mismatch contours J varying}.  As before, increasing the precession frequency $\tilde{\Omega}$ barely changes the shape of the contours, but it does increase the mismatches as a greater difference between the GW phase of the RP and NP waveforms accumulates during the inspiral.

A key difference between Figs~\ref{fig: opt mismatch contours J varying} and \ref{fig: opt mismatch contours N varying} is the presence in the latter of true nulls at which $F_+ = F_\times = 0$.  For the single L-shaped GW detector we consider, there are four true nulls located where the dot-dashed red and dotted purple curves intersect at ($\theta_S, \phi_S$) = ($\pi/2, (2n+1)\pi/4$) for integer $n$.  At these blind spots, the GW detector is insensitive to both GW polarizations ($\rho = 0$) and no mismatch $\epsilon$ can be large enough to yield detectable precession $\mathcal{I}_P > 0$.  In Fig.~\ref{fig: opt mismatch contours N varying}, the four true nulls are marked by red dots.  The smaller precession amplitude $\tilde{\theta} = 4$ in the top panels implies that the rings about the +~nulls are complete for sufficiently large precession frequency $\tilde{\Omega}$.  In the lower panels, the larger precession amplitude $\tilde{\theta} = 8$ implies that the true nulls have taken four bites shown in dark blue out of our two donuts denoting detectable precession.

\section{Detecting precession as a function of BBH masses, spins, and redshift}
\label{S:BBHpops}

Our study so far has fixed the chirp mass $\mathcal{M}$, mass ratio $q$, and redshift $z$ of our sources to focus the detectability of precession as a function of binary orientation $\hat{\Vec{J}}$ and sky location $\hat{\Vec{N}}$.  This leads to the dramatic contour plots shown in Figs~\ref{fig: SNR J and N} through \ref{fig: opt mismatch contours N varying}.  However, GW astronomers studying BBH populations will mostly be interested in sources with isotropic orientations distributed isotropically on the sky, but with a range of chirp masses, mass ratios, and redshifts.  We now present the fraction of systems in which precession is detectable for such isotropic distributions and varying values of these parameters in Tables~\ref{tab: MC on source and sky loc - fixed prec. param} and \ref{tab: MC on source and sky loc - MC prec. param}.  Unlike in Figs.~\ref{fig: opt mismatch contours J varying} through \ref{fig: opt mismatch contours N varying} in which we conservatively minimized $\epsilon_P$ and $\mathcal{I}_P$ over the nuisance parameter $\gamma_P$, in these tables we assume that it has its true flat distribution between 0 and $2\pi$ and use the Lindblom criterion of Eq.~(\ref{eq: lindblom criterion}) directly.

Consider first Table~\ref{tab: MC on source and sky loc - fixed prec. param} in which we use the precession amplitude $\tilde{\theta}$ and frequency $\tilde{\Omega}$ as independent parameters.  This showcases one of the advantages of the precessional parameter approach first proposed in \citeauthor{Ourpaper}~\cite{Ourpaper}; it allows us to disentangle the question of how precession arises from the detectability of that precession.  The first three rows of data show binary populations that only differ in their chirp mass $\mathcal{M}$.  Binaries with larger $\mathcal{M}$ have larger SNR because the GW amplitude $A \propto \mathcal{M}^{5/6}$ by Eq.~(\ref{eq:takahasi amplitude}), but they have shorter waveforms and thus fewer precession cycles $N_{\rm pre}$ in band according to Eq.~(\ref{E:Npre_est}), leading to smaller mismatches.  These two competing affects roughly cancel in the Lindblom criterion (\ref{eq: lindblom criterion}) for the $\mathcal{M} = 10 M_\odot$ and $20 M_\odot$ populations at $z = 0.1$, but the $\mathcal{M} = 40 M_\odot$ population has a small fraction of systems with detectable precession even at this low redshift.  As $z$ increases, all the populations have reduced SNR because of the higher luminosity distance $D_L$, but the populations with higher $\mathcal{M}$ have fractions with detectable precession that fall off even more rapidly because of the shorter waveforms ($f_{\rm cut} \propto \mathcal{M}^{-1}$).

Now consider the first, fourth, and fifth rows of data in Table~\ref{tab: MC on source and sky loc - fixed prec. param} whose populations differ only in their mass ratio $q$.  The fractions with detectable precession are only modestly smaller for the population with $q = 0.5$ compared to the default case with $q = 1$.  This follows because both $N_{\rm pre} \propto \eta^{-2/5}$ and $f_{\rm cut} \propto \eta^{3/5}$ only depend on $q$ through the symmetric mass ratio $\eta \equiv q/(1+q)^2$ that only decreases from $0.25$ to $0.222$ as $q$ drops from $1$ to $0.5$.  The fraction of systems with detectable precession drops dramatically for $q = 0.1$ for which $\eta = 0.083$.

Finally, we compare the default case shown in the first row of data in Table~\ref{tab: MC on source and sky loc - fixed prec. param} with populations with different values of precession amplitude $\tilde{\theta}$ and frequency $\tilde{\Omega}$ in the seventh through tenth rows.  The default values $\tilde{\theta} = 4$ and $\tilde{\Omega} = 2$ were chosen because they roughly correspond to the medians for an equal-mass BBH population with isotropically oriented maximal spins.  The lower (upper) values $\tilde{\theta} = 1$ and $\tilde{\Omega} = 1$ ($\tilde{\theta} = 8$ and $\tilde{\Omega} = 3$) were chosen because they roughly correspond to the $5^{\rm th}$ ($95^{\rm th}$) percentiles of this same population.  We see that precession amplitude has the largest impact on the fractions with detectable precession.  For 
$\tilde{\theta} = 1$ less than 10\% of systems have detectable precession even at $z = 0.2$, while for $\tilde{\theta} = 8$ there is still a greater than 10\% chance of detecting precession at $z = 0.75$.  The precession frequency $\tilde{\Omega}$ also has a significant effect on the detectability of precession since 
$N_{\rm pre} \propto \tilde{\Omega}$.  Although precession is detectable for greater than $90\%$ of systems at $z = 0.1$ for all values of $\tilde{\Omega}$, higher values of $\tilde{\Omega}$ allow a detectable number of precession cycles even at modest redshift.  For example, the detectable fractions at $z = 0.5$ are 2.2\%, 15.4\%, and 29.2\% for $\tilde{\Omega} = 1$, $2$, and $3$ respectively for $\mathcal{M} = 20 M_\odot$ and $q=1$.  Precession is virtually undetectable beyond $z = 1$ for all scenarios.

\begin{table*}
    \centering
    \begin{tabular}{|c|c|c|c|c|c|c|c|c|c|c|c|c|c|c|c|c|c|}
    \hline
    $\mathcal{M}$ & $q$ & $\chi_1$ & $\chi_2$ & Spin alignment & \multicolumn{3}{c|}{$\tilde{\theta}$} & \multicolumn{3}{c|}{$\tilde{\Omega}$} & $z = 0.1$ & $0.2$ & $0.3$ & $0.4$ & $0.5$ & $0.75$ & $1$\\
    \hline
    \hline
    20 $M_{\odot}$& 1 & 1 & 1 & ISO & 0.96 & 4.20 & 7.64 & 1.61 & 2.10 & 2.35 & 92.0 & 75.9 & 57.4 & 38.2 & 22.1 & 2.6 & 0.1 \\
    \hline
    10 $M_{\odot}$& 1 & 1 & 1 & ISO & 1.01 & 4.25 & 7.51 & 1.70 & 2.07 & 2.29 & 92.0 & 78.9 & 63.7 & 48.9 & 35.8 & 12.0 & 2.5 \\
    \hline
    40 $M_{\odot}$& 1 & 1 & 1 & ISO & 1.00 & 4.21 & 7.72 & 1.53 & 2.15 & 2.42 & 76.8 & 40.3 & 13.5 & 2.4 & 0.1 & 0.0 & 0.0 \\
    \hline
    20 $M_{\odot}$& 0.95 & 1 & 1 & ISO & 1.33 & 4.25 & 7.59 & 1.62 & 2.10 & 2.35 & 93.8 & 77.5 & 58.1 & 38.6 & 22.2 & 2.7 & 0.1 \\
    \hline
    20 $M_{\odot}$& 0.5 & 1 & 1 & ISO & 2.73 & 6.10 & 8.53 & 0.95 & 1.55 & 2.38 & 96.9 & 85.3 & 64.6 & 39.7 & 19.8 & 1.2 & 0.0 \\
    \hline
    20 $M_{\odot}$& 0.1 & 1 & 1 & ISO & 1.51 & 5.59 & 11.06 & 1.05 & 1.28 & 1.37 & 85.5 & 53.0 & 19.7 & 3.7 & 0.2 & 0.0 & 0.0 \\
    \hline
    10 $M_{\odot}$& 0.1 & 1 & 1 & ISO & 1.83 & 6.65 & 13.51 & 0.73 & 1.06 & 1.17 & 94.3 & 81.5 & 63.9 & 42.6 & 22.1 & 1.2 & 0.0 \\
    \hline
    20 $M_{\odot}$& 1 & 0.5 & 0.5 & ISO & 0.50 & 2.15 & 3.75 & 1.82 & 2.03 & 2.19 & 79.1 & 48.6 & 24.8 & 9.9 & 3.1 & 0.0 & 0.0 \\
    \hline
    20 $M_{\odot}$& 1 & 0.2 & 0.2 & ISO & 0.20 & 0.86 & 1.48 & 1.94 & 2.02 & 2.09 & 42.2 & 8.7 & 1.0 & 0.1 & 0.0 & 0.0 & 0.0 \\
    \hline
    20 $M_{\odot}$& 1 & 0.1 & 0.1 & ISO & 0.10 & 0.44 & 0.75 & 1.98 & 2.01 & 2.05 & 11.4 & 0.3 & 0.0 & 0.0 & 0.0 & 0.0 & 0.0 \\
    \hline
    20 $M_{\odot}$& 0.95 & 1 & 1 & WA & 0.34 & 1.27 & 2.31 & 2.36 & 2.37 & 2.38 & 65.1 & 27.0 & 8.3 & 1.8 & 0.2 & 0.0 & 0.0 \\
    \hline
    20 $M_{\odot}$& 0.95 & 0.5 & 0.5 & WA & 0.24 & 0.74 & 1.35 & 2.23 & 2.24 & 2.24 & 38.5 & 6.7 & 0.7 & 0.0 & 0.0 & 0.0 & 0.0 \\
    \hline
    20 $M_{\odot}$& 0.95 & 0.2 & 0.2 & WA & 0.15 & 0.34 & 0.58 & 1.95 & 2.12 & 2.13 & 5.2 & 0.0 & 0.0 & 0.0 & 0.0 & 0.0 & 0.0 \\
    \hline
    20 $M_{\odot}$& 0.95 & 1 & 1 & SA & 0.12 & 0.43 & 0.77 & 2.37 & 2.37 & 2.37 & 13.6 & 0.5 & 0.0 & 0.0 & 0.0 & 0.0 & 0.0 \\
    \hline
    20 $M_{\odot}$& 0.95 & 0.5 & 0.5 & SA & 0.08 & 0.25 & 0.46 & 2.24 & 2.24 & 2.25 & 2.0 & 0.0 & 0.0 & 0.0 & 0.0 & 0.0 & 0.0 \\
    \hline
    20 $M_{\odot}$& 0.95 & 0.2 & 0.2 & SA & 0.05 & 0.12 & 0.20 & 1.95 & 2.13 & 2.13 & 0.0 & 0.0 & 0.0 & 0.0 & 0.0 & 0.0 & 0.0 \\
    \hline
    20 $M_{\odot}$& 0.5 & 1 & 0.1 & ISO & 2.13 & 6.23 & 7.61 & 0.94 & 1.57 & 1.81 & 95.9 & 84.0 & 63.8 & 39.4 & 19.2 & 0.9 & 0.0 \\
    \hline
    20 $M_{\odot}$& 0.5 & 0.1 & 1 & ISO & 0.75 & 1.62 & 1.90 & 1.49 & 2.57 & 2.66 & 77.8 & 34.5 & 8.9 & 1.4 & 0.0 & 0.0 & 0.0 \\
    \hline
    \end{tabular}
    \caption{Percentage of binaries with detectable precession from isotropically oriented (in $\hat{\Vec{N}}$ and $\hat{\Vec{J}}$) populations with different chirp masses $\mathcal{M}$, mass ratios $q$, and dimensionless spins $\chi_1$ and $\chi_2$ indicated by the values in the first four columns.  The fifth column indicates the spin alignment of the population: ISO - isotropic, WA (SA) - weakly (strongly) aligned, i.e. spins within a cone with an opening angle of $30^\circ$ ($10^\circ$) about the orbital angular momentum as $r \to \infty$.  The sixth through eighth (ninth through eleventh) columns show the $5^{\rm th}$, $50^{\rm th}$, and $95^{\rm th}$ percentiles of the distributions of the precession amplitude $\tilde{\theta}$ (precession frequency $\tilde{\Omega}$) at $f = f_{\rm min}$ resulting from these BBH mass and spin distributions.  The remaining columns show the percentages of systems with detectable precession for these populations at different redshifts $z$ with values indicated in the top row.  We use $10^4$ binaries for each row in this table leading to $\sim 1\%$-level errors due to Poisson noise.
    } \label{tab: MC on source and sky loc - MC prec. param}
\end{table*}

In Table~\ref{tab: MC on source and sky loc - MC prec. param}, we consider binary populations in which the precession parameters $\tilde{\theta}$ and $\tilde{\Omega}$ are not held constant but instead are calculated from underlying distributions of misaligned BBH spins. We calculate these precession parameters using the precession average of Eq.~(\ref{eq:precession average}) as implemented in the \texttt{precession} \cite{NewPrecessionCode} package. Our default case, shown in the first row of data, again has $\mathcal{M} = 20 M_\odot$ and $q = 1$ as in the first row of Table~\ref{tab: MC on source and sky loc - fixed prec. param}.  However, we allow the precessional parameters to have the natural distribution arising from a binary population with isotropic, maximal spins.  This yields reasonably broad distributions of the precessional parameters with median values $\tilde{\theta} = 4.2$ and $\tilde{\Omega} = 2.1$ quite close to the constant values $\tilde{\theta} = 4$ and $\tilde{\Omega} = 2$ in the first row of Table~\ref{tab: MC on source and sky loc - fixed prec. param}.  Comparing these two populations, we see that the broader distribution of precessional parameters in Table~\ref{tab: MC on source and sky loc - MC prec. param} implies a lower fraction of systems with detectable precession for $z \leq 0.3$ but a higher fraction at $z \geq 0.4$.

In the first three rows of data in Table~\ref{tab: MC on source and sky loc - MC prec. param}, we again vary the chirp mass $\mathcal{M}$ while keeping the other parameters constant.  We see that the distributions of the precessional parameters are very similar.  This is expected because general relativity is a scale-free theory and the scaling of the dimensionless parameters in Eq.~(\ref{E:PPfreqdep}) was chosen to make them weakly sensitive to the reference frequency at which they are normalized (see Appendix~\ref{Appendix: PN validations} for more details on this normalization).  These populations therefore exhibit the same trends with mass as we saw in Table~\ref{tab: MC on source and sky loc - fixed prec. param}.

In the fourth through sixth rows of data in Table~\ref{tab: MC on source and sky loc - MC prec. param}, we again vary the mass ratio $q$ while keeping the other parameters constant.  This does change the distributions of precessional parameters, increasing the median and variance of $\tilde{\theta}$ while decreasing the median and variance of $\tilde{\Omega}$.  These two changes partially offset, but comparing rows with the same $\mathcal{M}$ and $q$ in Tables~\ref{tab: MC on source and sky loc - fixed prec. param} and \ref{tab: MC on source and sky loc - MC prec. param} shows that the latter have higher fractions with detectable precession at $z \geq 0.4$, particularly for the most extreme mass ratio $q = 0.1$.

In the eighth, ninth, and tenth rows of data in Table~\ref{tab: MC on source and sky loc - MC prec. param}, we examine how BBH spin magnitude affects the precession parameters and consequent detectability of precession as a function of redshift.  As expected from the top panel of Fig.~\ref{fig: orbital frame}, the precession amplitude $\tilde{\theta}$ is linearly proportional to the BBH spin magnitudes in the limit $S \ll L$.  The median precession frequency $\tilde{\Omega}$ is nearly independent of spin magnitude, while its standard deviation is roughly linear in the spin magnitude.  The reduction in $\tilde{\theta}$ makes precession more difficult to detect, particularly at high redshift.  An analysis of the Gravitational Wave Transient Catalog 3 (GWTC-3) indicates that the observed median spin magnitude is $\chi_i \approx 0.26$ \cite{GWTCPOP}.  If this claim is accurate, precession will be extremely challenging to observe beyond $z \gtrsim 0.1$.

In the eleventh through sixteenth rows of data in Table~\ref{tab: MC on source and sky loc - MC prec. param}, we investigate how BBH spin alignment affect the precession parameters and the detectability of precession.  We consider weakly aligned (WA) and strongly aligned (SA) systems in which the individual BBH spins are aligned within $30^\circ$ and $10^\circ$ respectively of the the orbital angular momentum at large binary separations.  Although astrophysical formation scenarios are beyond the scope of this work, such partial alignment might arise in BBHs formed from stellar binaries or through interaction with supermassive black hole accretion disks.  We adopt a mass ratio $q = 0.95$ for these systems rather than unity because the constancy of the total spin magnitude $S$ for precisely equal-mass systems prevents the standard precession-averaged evolution of these systems as implemented in the \texttt{precession} \cite{NewPrecessionCode} package.

Unsurprisingly, spin alignment with the orbital angular momentum leads to alignment between the orbital and total angular momenta, and thus to smaller precession amplitudes.  Assuming that the precession amplitude $\tilde{\theta}$ scales linearly with the individual spins $\chi_i$ for isotropic systems, WA (SA) reduces the median value of $\tilde{\theta}$ by the same amount as decreasing the individual spin magnitudes from maximal to $\chi_i \approx 0.3$ ($\chi_i \approx 0.1$).  This alignment also increases the median and reduces the variance of the precession frequency $\tilde{\Omega}$, but this cannot compensate for the reduction in $\tilde{\theta}$.  Only with maximal spins can precession be detected in a majority of WA systems, even in our lowest redshift bin $z = 0.1$.  Precession is rarely detected in SA systems in any of the populations we considered.

In the final two rows of Table~\ref{tab: MC on source and sky loc - MC prec. param}, we investigate the effects of asymmetries between the two BBH spins in unequal-mass systems ($q = 0.5$).  We see that, compared to systems with the same mass ratio and two maximal spins shown in the fifth row of data, a reduced secondary spin has a small effect on both the precession parameters and the fraction of systems with detectable precession as a function of redshift.  Reducing the primary spin significantly reduces the precession amplitude $\tilde{\theta}$.  However, it also increases the median precession frequency $\tilde{\Omega}$ to the highest value of all the populations considered in this table.  This is expected for a population with just such a spin asymmetry (see Eq.~(10) of \cite{Ourpaper}).  This high precession frequency does not fully compensate for the reduced precession amplitude, but significant fractions of these systems still exhibit detectable precession at $z \lesssim 0.2$.

\section{Discussion} 
\label{sec: conclusions}

In a generic BBH system, each black hole will have a spin $\Vec{S}_i$ that is misaligned with both the other black-hole spin and the orbital angular momentum $\Vec{L}$ of the binary.  Spin-orbit and spin-spin coupling sourced by these misaligned spins will cause $\Vec{L}$ to precess and nutate about the total angular momentum $\Vec{J} = \Vec{L} + \vec{S}_1 + \vec{S}_2$ which is conserved on the precession timescale.  To lowest PN order, the GWs emitted by the binary are preferentially beamed in the direction of $\Vec{L}$ according to Eq.~(\ref{eq: GWstrains}).  This implies that precession and nutation will modulate the GW amplitude and phase.

Failing to properly account for these modulations will introduce errors into the parameter estimates for individual BBH systems and biases into studies of BBH populations \cite{Chandramouli2025}. In our previous study~\cite{Ourpaper}, we introduced a taxonomy of BBH precession and nutation and defined five new phenomenological parameters to describe generic (simple) precession that evolve on the radiation-reaction timescale. 

In this paper, we focus on regular precession in which only two of the five precessional parameters are nonzero: the precession amplitude $\langle \theta_{LJ} \rangle$ and precession frequency $\langle \Omega_{LJ} \rangle$.  Although in reality these parameters are functions of the individual BBH masses and spins and evolve on the radiation-reaction timescale, we assume that their lowest-order PN dependence on these parameters persists throughout the inspiral.  We use this assumption to define dimensionless versions of these parameters $\tilde{\theta}$ and $\tilde{\Omega}$ in Eq.~(\ref{E:PPfreqdep}) that are constant throughout the inspiral.  We then use these parameters to define a toy model of a regularly precessing waveform as described in Sec.~\ref{sec: RP model}.  This toy model includes $\gamma_P$, the integration constant to Eq.~(\ref{eq: PhiLJ}), as an additional parameter.  As $\gamma_P$ is the azimuthal angle that defines the direction of the orbital angular momentum $\Vec{L}$ with respect to the total angular momentum $\Vec{J}$, it is an extrinsic parameter with a flat distribution between $0$ and $2\pi$.  We treat it as a nuisance parameter in most of our analysis.

In Figs.~\ref{fig: Amplitude various p} and \ref{fig: phi_p plus 2 deltaphi p}, we examined how our three parameters $\tilde{\theta}$, $\tilde{\Omega}$, and $\gamma_P$ affect GW amplitudes and phases for three different binary orientations (face-on, edge-on, and generic) listed in Table~\ref{tab: source parameters}.  We then defined $\epsilon_P$, the mismatch minimized over $\gamma_P$, in Eq.~(\ref{eq: mismatch gamma_p}) and showed in Fig.~\ref{fig: mismatch vs omega} that it is a conservative estimate of the difference between regularly precessing (RP) and non-precessing (NP) waveforms.  Fig.~\ref{fig: opt mismatch for all systems} shows contour plots of this newly defined $\epsilon_P$ for the three binary orientations listed in Table~\ref{tab: source parameters}, three choices of chirp mass $\mathcal{M}$, and three different mass ratios $q$.  We then use $\epsilon_P$ to define a new quantity $\mathcal{I}_P$ in Eq.~(\ref{eq: lindblom criterion modified}) that according to the Lindblom criterion \cite{Lindblom} will be positive for RP systems that can be distinguished from NP templates, i.e. precession is detectable in these systems.  Figs.~\ref{fig: opt mismatch contours J varying} and \ref{fig: opt mismatch contours N varying} show contours of $\mathcal{I}_P$ for different binary orientations and sky locations respectively.  They reveal that precession is most detectable in systems for which the orbital angular momentum $\Vec{L}$ precesses through configurations in which $\hat{\Vec{L}} \cdot \hat{\Vec{N}} = F_+ = 0$ during the inspiral.  Precession is least detectable in systems with binary orientations for which $\hat{\Vec{J}} \cdot \hat{\Vec{N}} = F_\times = 0$ and systems located at the true nulls ($\theta_S, \phi_S$) = ($\pi/2, (2n+1)\pi/4$) where $F_+ = F_\times = 0$.

In Sec.~\ref{S:BBHpops}, we calculated the fraction of systems in which precession is detectable for different BBH populations.  The populations listed in Table~\ref{tab: MC on source and sky loc - fixed prec. param} are characterized by our newly defined precessional parameters $\tilde{\theta}$ and $\tilde{\Omega}$, while we calculated the distributions of these parameters from underlying BBH mass and spin distributions in Table~\ref{tab: MC on source and sky loc - MC prec. param}.  Note that these populations were chosen to explore the allowable BBH parameter space and are not intended to represent astrophysical BBH populations produced in a particular formation channel as would be generated by a self-consistent population-synthesis code.  Populations for which precession was detectable in a high fraction of systems generally had both large precession amplitudes and a high number of precession cycles $N_{\rm pre}$~(\ref{E:Npre_est}) within the GW sensitivity band.  Such populations had large values of the precessional parameters $\tilde{\theta}$ and $\tilde{\Omega}$.  Table~\ref{tab: MC on source and sky loc - MC prec. param} revealed that large values of $\tilde{\theta}$ are found in populations with large spin magnitudes and significant fractions of misaligned spins as are found for isotropic spin orientations.  Precession frequencies $\tilde{\Omega}$ are less sensitive to the underlying BBH mass and spin distributions; the ratio of the maximum to the minimum value of the median $\tilde{\Omega}$ between populations in Table~\ref{tab: MC on source and sky loc - MC prec. param} was only $\approx 2.4$ while this same ratio was $\approx 55$ for $\tilde{\theta}$.  As predicted in Sec.~IIC of our previous study \cite{Ourpaper}, the precession frequency is maximized (minimized) in populations with small mass ratios and spin magnitudes such that $S_2 > S_1$ ($S_2 < S_1$).

Our analysis of the detectability of precession was carried out for a single L-shaped detector with the noise power spectral density $S_n(f)$ of a single advanced LIGO detector at design sensitivity \cite{PhysRevD.46.5236,PhysRevD.46.1517}.
A detector network would yield higher fractions of systems with detectable precession, both because of the larger network SNRs and the inclusion of additional +~nulls (at which precession is most easily detectable) for detectors with different orientations. The detectors at the two LIGO sites at Livingston, Louisiana and Hanford, Washington were chosen oriented at a $90^\circ$ angle with respect to each other to maximize their correlated response to a GW signal and thus reduce the effects of uncorrelated noise.  Unfortunately, this conservative choice yielded the same polarization sensitivity for the two detectors and +~nulls located at the same binary orientations and sky locations, rather than distributed more evenly through parameter space.  Nonetheless, an analysis including all of the detectors in the LVK network would still yield higher fractions of detectable precession than that presented in this paper.  

Even tighter constraints on precession will be provided by future third-generation GW detectors like the Einstein Telescope \cite{ET} and Cosmic Explorer \cite{CE}.  These detectors have lower noise power spectral densities $S_n(f)$, particularly at low GW frequencies $f$, leading to higher SNRs $\rho$ and larger mismatches $\epsilon$ due to the greater number of precession cycles in band.  Both of these effects increase $\mathcal{I}_P$ (\ref{eq: lindblom criterion modified}) and thus the detectability of precession according to the Lindblom criterion.   

In future work, we will generalize our analysis from regular precession to generic (simple) precession \cite{GenericPrecession}.  The latter includes nutation, i.e. oscillation of the angle $\theta_{LJ}$ between the orbital and total angular momenta on the precession timescale.  According to our previous work (reviewed in Sec.~\ref{subs: taxonomy} above), this nutation can be characterized by a nutation amplitude $\Delta\theta_{LJ}$, nutation frequency $\omega$, and integration constant $\gamma_N$ giving the nutation phase at a GW frequency $f_{\rm min}$.  In addition, for generic precession, the precession frequency oscillates at the nutation frequency and an additional parameter $\Delta\Omega_{LJ}$ is needed to characterize the amplitude of these oscillations.  We expect nutation to be even more challenging to detect than precession for two reasons: (1) the oscillatory nature of nutation does not generate a secular contribution to the GW phase, and (2) nutation amplitudes are at most $\approx 20\%$ as large as precession amplitudes.

An additional avenue of research would be to explore more sophisticated metrics than the Lindblom criterion of Eq.~(\ref{eq: lindblom criterion}) \cite{Lindblom}.  The evidence ratio or Bayes' factor between regularly precessing and non-precessing template families would give a more accurate prediction of whether a given source is precessing.  Such an approach would account for the higher dimensionality of the parameter space of RP waveforms, as well as possible degeneracies between precessing and non-precessing parameters.  Preliminary studies suggest that such degeneracies are small since none of the non-precessing parameters can induce oscillations of the GW amplitude and phase like those seen in Figs.~\ref{fig: Amplitude various p} and \ref{fig: phi_p plus 2 deltaphi p}.  Although large residual eccentricities might produce oscillations in the waveform as the system oscillates between pericenter and apocenter, such oscillations would occur on the much shorter orbital timescale and should be distinguishable in systems spending at least a full precession cycle in band.  It would also be valuable to reanalyze our results on the detectability of precession using the precession signal-to-noise ratio $\rho_{\rm p}$ introduced by \citeauthor{Fairhurst2020rho_p}~\cite{Fairhurst2020rho_p} and further developed in subsequent work \cite{Fairhurst2020twoharmonics, Hoy2024Precession}.

The next step beyond detecting precession (a model-selection problem) would be estimating the values of the precession parameters within the context of a given precessing waveform model.  Such parameter estimates would indirectly constrain BBH spins and thus astrophysical formation channels for BBH systems. 
Dynamical interactions within dense stellar clusters generally produce BBH systems with isotropically oriented spins \cite{Benacquista2013,Rodriguez2016}.  Isolated stellar binaries evolve into BBHs whose spins are mostly aligned with their orbital angular momentum, although natal kicks at birth may provide a source of some misalignment \cite{Kalogera2000,Postnov2014,GerosaFishbach2021,SteinleKesden2022}.  A range of possible spin misalignments are predicted in other scenarios such as BBHs embedded in AGN disks or formed from primordial black holes. 
Our results shown in Table~\ref{tab: MC on source and sky loc - MC prec. param} suggest that the LVK network at its current sensitivity has the potential to identify regularly precessing binaries in a wide range of scenarios, at least at low redshift.  This further supports the contention that observing precession may be the key to revealing the origins of stellar-mass BBHs.

\acknowledgements

The authors thank Tien Nguyen, Christina McNally, and Floor Broekgaarden for insightful discussions.

T.S., M.K., and L.K. are supported by the National Science Foundation Grant No. PHY-2309320.
E.S., S.A., M.K., and L.K. were supported by National Science Foundation Grant No. PHY-2011977.  N.S. is supported by the Leverhulme Trust Grant No. RPG-2019-350 and the Natural Sciences and Engineering Research Council of Canada (NSERC) through the Canada Research Chairs and Discovery Grants programs.  The authors acknowledge the Texas Advanced Computing Center (TACC) at The University of Texas at Austin for providing HPC resources that have contributed to the research results reported within this paper \cite{10.1145/3093338.3093385}. URL: \href{http://www.tacc.utexas.edu}{http://www.tacc.utexas.edu}

\textit{Software:} This work made use of the following software packages: \texttt{astropy} \citep{astropy:2013, astropy:2018, astropy:2022}, 
\texttt{matplotlib} \citep{Hunter:2007}, \texttt{numpy} \citep{numpy}, \texttt{scipy} \citep{2020SciPy-NMeth, scipy_10909890}, \texttt{PyCBC} \citep{alex_nitz_2022_6324278}, and \texttt{precession} \citep{Gerosa_Precession_2016, NewPrecessionCode}.  Software citation information aggregated using \texttt{\href{https://www.tomwagg.com/software-citation-station/}{The Software Citation Station}} \citep{software-citation-station-paper, software-citation-station-zenodo}.

\appendix

\section{Three reference frames}
\label{Appendix: 3frames}

We use three distinct reference frames to calculate our regularly precessing waveforms.

\subsection{Detector frame}

The \emph{detector frame} is chosen such that its basis vectors $\hat{\Vec{X}}_D$ and $\hat{\Vec{Y}}_D$ point along the arms of the interferometer, and $\hat{\Vec{Z}}_D = \hat{\Vec{X}}_D \times \hat{\Vec{Y}}_D$ points towards the zenith at the detector location.  The sky location of the binary $\hat{\Vec{N}}$ and the direction of the total angular momentum $\hat{\Vec{J}}$ are specified respectively by the angles $(\theta_S, \Phi_S)$ and $(\theta_J, \Phi_J)$ with respect to this frame:
\begin{subequations}
\begin{align}
\hat{\Vec{N}} &= \sin\theta_S(\cos\phi_S\hat{\Vec{X}}_D + \sin\phi_S\hat{\Vec{Y}}_D) + \cos\theta_S\hat{\Vec{Z}}_D\,, \\
\hat{\Vec{J}} &= \sin\theta_J(\cos\phi_J\hat{\Vec{X}}_D + \sin\phi_J\hat{\Vec{Y}}_D) + \cos\theta_J\hat{\Vec{Z}}_D\,.
\end{align}
\end{subequations}

\subsection{Sky frame}

The \emph{sky frame} is chosen such that $\hat{\Vec{N}}$ points into the plane of the sky (opposite to the direction of GW propagation), $\hat{\Vec{V}}$ points towards the zenith, and $\hat{\Vec{H}} = \hat{\Vec{V}} \times \hat{\Vec{N}}$ is parallel to the horizon:
\begin{subequations}
\begin{align}
\hat{\Vec{H}} &= -\sin\phi_S\hat{\Vec{X}}_D + \cos\phi_S\hat{\Vec{Y}}_D\,, \\
\hat{\Vec{V}} &= -\cos\theta_S(\cos\phi_S\hat{\Vec{X}}_D + \sin\phi_S\hat{\Vec{Y}}_D) + \sin\theta_S\hat{\Vec{Z}}_D\,.
\end{align}
\end{subequations}

\subsection{Source frame}

The \emph{source frame} is chosen such that $\hat{\Vec{Z}_S}$ points along the direction of the total angular momentum $\hat{\Vec{J}}$,
\begin{equation}
\hat{\Vec{X}}_S = \frac{\hat{\Vec{N}} \times \hat{\Vec{J}}}{|\hat{\Vec{N}} \times \hat{\Vec{J}}|}
\end{equation}
points along the line of ascending node of the plane perpendicular to $\Vec{J}$ with respect to the sky plane spanned by $\hat{\Vec{H}}$ and $\hat{\Vec{V}}$, and $\hat{\Vec{Y}}_S = \hat{\Vec{Z}}_S \times \hat{\Vec{X}}_S$.  We define the inclination $\iota_{JN}$ as the angle between $\hat{\Vec{J}}$ and $\hat{\Vec{N}}$,
\begin{align} 
\cos\iota_{JN} &= \hat{\Vec{J}} \cdot \hat{\Vec{N}} \label{eq: cosiota} \\
&= \sin\theta_J\sin\theta_S\cos(\phi_J - \phi_S) + \cos\theta_J\cos\theta_S \,, \notag 
\end{align}
and the longitude of ascending node $\Omega_{XH}$ such that the source and sky frames are related by
\begin{subequations}
\begin{align}
\hat{\Vec{X}}_S &= \cos\Omega_{XH} \hat{\Vec{H}} + \sin\Omega_{XH} \hat{\Vec{V}}\,, \\
\hat{\mathbf{Y}}_S &= \cos\iota_{JN}(-\sin\Omega_{XH}\hat{\mathbf{H}} + \cos\Omega_{XH}\hat{\mathbf{V}}) + \sin\iota_{JN}\hat{\mathbf{N}}\,, \\
\hat{\mathbf{Z}}_S &= \sin\iota_{JN}(\sin\Omega_{XH}\hat{\mathbf{H}} - \cos\Omega_{XH}\hat{\mathbf{V}})+\cos\iota_{JN}\hat{\mathbf{N}}\,,
\end{align}
\end{subequations}
where
\begin{subequations} \label{eq: LoAN}
\begin{align}
\sin\Omega_{XH} &= \frac{\sin\theta_J\sin(\phi_J - \phi_S)}{\sin\iota_{JN}} \,, \\
\cos\Omega_{XH} &= \frac{\sin\theta_J\cos\theta_S\cos(\phi_J - \phi_S) - \cos\theta_J\sin\phi_S}{\sin\iota_{JN}} \,.
\end{align}
\end{subequations}
In the aligned case ($\iota_{JN} = 0$), Eq.~(\ref{eq: LoAN}) breaks down and one can choose $\Omega_{XH} = 0$ without loss of generality (it is degenerate with the integration constant $\gamma_P$).

In the source and sky frames, the direction of the orbital angular momentum $\hat{\Vec{L}}$ is given by

\begin{widetext}
\begin{align}
\hat{\Vec{L}} &= \sin\theta_{LJ}[\cos\Phi_{LJ} \hat{\Vec{X}}_S + \sin\Phi_{LJ} \hat{\Vec{Y}}_S] + \cos\theta_{LJ} \hat{\Vec{Z}}_S \,, \\
&= [\sin\theta_{LJ}(\cos\Phi_{LJ}\cos\Omega_{XH} - \sin\Phi_{LJ}\cos\iota_{JN}\sin\Omega_{XH})
+ \sin\iota_{JN}\sin\Omega_{XH}\cos\theta_{LJ}]\hat{\mathbf{H}} \notag \\
&\quad +[\sin\theta_{LJ}(\cos\Phi_{LJ}\sin\Omega_{XH} + \sin\Phi_{LJ}\cos\iota_{JN}\cos\Omega_{XH})
- \sin\iota_{JN}\cos\Omega_{XH}\cos\theta_{LJ}]\hat{\mathbf{V}}
\notag \\
&\quad +(\sin\theta_{LJ}\sin\Phi_{LJ}\sin\iota_{JN} + \cos\theta_{LJ}\cos\iota_{JN})\hat{\mathbf{N}}\,.
\end{align}
\end{widetext}
The source and sky frames are depicted in green and red respectively in Fig.~\ref{fig: orbital frame}.  In our regularly precessing waveform model, only the angles $\Phi_{LJ}$ and $\theta_{LJ}$ are time dependent, and they vary on the precession and radiation-reaction timescales respectively according to Eqs.~(\ref{eq: PhiLJ}) and (\ref{eq: thetaLJ}).

\subsection{GW strain}

The observed GW strain given by Eq.~(\ref{eq: hf}) depends on the relative orientation of these three frames through the GW fields $h_+$ and $h_\times$ of Eq.~(\ref{eq: GWstrains}) and the detector beam-pattern coefficients $F_+$ and $F_\times$ of Eq.~(\ref{eq: BeamPatt}).  The latter depend on the polarization angle $\psi$ of Eq.~(\ref{eq: polarization angle psi}) which is given in terms of the above angles as

\begin{widetext}
\begin{align}
\hat{\Vec{L}} \cdot \hat{\Vec{Z}}_D &= \sin\theta_{LJ}[\cos\Phi_{LJ}\sin\Omega_{XH}\sin\theta_S
+ \sin\Phi_{LJ}(\cos\iota_{JN}\cos\Omega_{XH}\sin\theta_S
+ \sin\iota_{JN}\cos\theta_S)] \notag \\
&\quad + \cos\theta_{LJ}(\cos\iota_{JN}\cos\theta_S
- \sin\iota_{JN}\cos\Omega_{XH}\sin\theta_S)\,, \\
\hat{\Vec{L}} \cdot \hat{\Vec{N}} &= \sin\theta_{LJ}\sin\iota_{JN}\sin\Phi_{LJ} + \cos\theta_{LJ}\cos\iota_{JN}\,, \label{eq:LN} \\
\hat{\Vec{N}} \cdot \hat{\Vec{Z}}_D &= \cos\theta_S\,, \\
\hat{\Vec{N}} \cdot (\hat{\Vec{L}} \times \hat{\Vec{Z}}_D) &= \hat{\Vec{Z}}_D \cdot (\hat{\Vec{N}} \times \hat{\Vec{L}}) \notag \\
&= [\sin\theta_{LJ}(\cos\Phi_{LJ}\cos\Omega_{XH}
- \sin\Phi_{LJ}\cos\iota_{JN}\sin\Omega_{XH})
+ \sin\iota_{JN}\sin\Omega_{XH}\cos\theta_{LJ}]\sin\theta_S\,, \\
\tan\psi &= \frac{\hat{\Vec{L}} \cdot \hat{\Vec{Z}}_D
- (\hat{\Vec{L}} \cdot \hat{\Vec{N}})(\hat{\Vec{N}} \cdot \hat{\Vec{Z}}_D)}{\hat{\Vec{N}} \cdot (\hat{\Vec{L}} \times \hat{\Vec{Z}}_D)} \\
 &= \frac{\sin\theta_{LJ}(\cos\Phi_{LJ}\sin\Omega_{XH}
+ \sin\Phi_{LJ}\cos\iota_{JN}\cos\Omega_{XH}) - \cos\theta_{LJ}\sin\iota_{JN}\cos\Omega_{XH}}{\sin\theta_{LJ}(\cos\Phi_{LJ}\cos\Omega_{XH} - \sin\Phi_{LJ}\cos\iota_{JN}\sin\Omega_{XH})
+ \cos\theta_{LJ}\sin\iota_{JN}\sin\Omega_{XH}} \label{eq: tanpsi}
\end{align}
\end{widetext}
In the aligned case ($\iota_{JN} = 0$), Eq.~(\ref{eq: tanpsi}) above reduces to $\psi = \Phi_{LJ} + \Omega_{XH}$,

and the longitude of ascending node $\Omega_{XH}$ is degenerate with the integration constant $\gamma_P$.  We can also use these three reference frames to evaluate the contribution $\delta\Phi$ to the GW phase given by Eq.~(\ref{eq: delta correction}):
\begin{widetext}
\begin{align}
\frac{\hat{d\Vec{L}}}{df} &= \langle\Omega_{LJ}\rangle \left( \frac{df}{dt} \right)^{-1} \sin\theta_{LJ}(-\sin\Phi_{LJ} \hat{\Vec{X}}_S + \cos\Phi_{LJ} \hat{\Vec{Y}}_S) + \frac{\langle\theta_{LJ}\rangle}{3f} \{ \cos\theta_{LJ}[\cos\Phi_{LJ} \hat{\Vec{X}}_S + \sin\Phi_{LJ} \hat{\Vec{Y}}_S] - \sin\theta_{LJ} \hat{\Vec{Z}}_S \} \\
(\hat{\Vec{L}} \times \hat{\Vec{N}}) \cdot \hat{\Vec{X}}_S &= (\hat{\Vec{X}}_S \times \hat{\Vec{L}}) \cdot \hat{\Vec{N}} = (\sin\theta_{LJ}\sin\Phi_{LJ} \hat{\Vec{Z}}_S - \cos\theta_{LJ} \hat{\Vec{Y}}_S) \cdot \hat{\Vec{N}} = \sin\theta_{LJ}\sin\Phi_{LJ} \cos\iota_{JN} - \cos\theta_{LJ} \sin\iota_{JN} \\
(\hat{\Vec{L}} \times \hat{\Vec{N}}) \cdot \hat{\Vec{Y}}_S &= (\hat{\Vec{Y}}_S \times \hat{\Vec{L}}) \cdot \hat{\Vec{N}} = (-\sin\theta_{LJ}\cos\Phi_{LJ} \hat{\Vec{Z}}_S + \cos\theta_{LJ} \hat{\Vec{X}}_S) \cdot \hat{\Vec{N}} = -\sin\theta_{LJ}\cos\Phi_{LJ} \cos\iota_{JN} \\
(\hat{\Vec{L}} \times \hat{\Vec{N}}) \cdot \hat{\Vec{Z}}_S &= (\hat{\Vec{Z}}_S \times \hat{\Vec{L}}) \cdot \hat{\Vec{N}} = \sin\theta_{LJ}(-\sin\Phi_{LJ}\hat{\Vec{X}}_S + \cos\Phi_{LJ} \hat{\Vec{Y}}_S) \cdot \hat{\Vec{N}} = \sin\theta_{LJ}\cos\Phi_{LJ} \sin\iota_{JN} \\
\frac{d\delta\Phi}{df} &= \left[ \frac{\hat{\Vec{L}} \cdot \hat{\Vec{N}}}{1 -(\hat{\Vec{L}} \cdot \hat{\Vec{N}})^{2}} \right] (\hat{\Vec{L}} \times \hat{\Vec{N}}) \cdot \frac{\hat{d\Vec{L}}}{df} \notag \\
&= \left[ \frac{\hat{\Vec{L}} \cdot \hat{\Vec{N}}}{1 -(\hat{\Vec{L}} \cdot \hat{\Vec{N}})^{2}} \right] \left[ \langle\Omega_{LJ}\rangle \left( \frac{df}{dt} \right)^{-1} \sin\theta_{LJ} (\cos\theta_{LJ}\sin\iota_{JN}\sin\Phi_{LJ} - \sin\theta_{LJ}\cos\iota_{JN}) - \frac{\langle\theta_{LJ}\rangle}{3f} \cos\Phi_{LJ} \sin\iota_{JN} \right] \label{E:ddeltaPhidf}
\end{align}
\end{widetext}

If the orbital angular momentum is aligned or anti-aligned with the line of sight at a frequency $f = f_{\rm al}$, we can Taylor expand Eq.~(\ref{E:ddeltaPhidf}) in powers of $\delta f \equiv f - f_{\rm al}$ about $f = f_{\rm al}$ to show that this equation remains finite.
\begin{widetext}
\begin{align}
\sin\theta_{LJ} &\approx \sin\iota_{JN} + \left( \frac{\langle\theta_{LJ}\rangle}{3f} \cos\iota_{JN} \right) \delta f - \frac{\langle\theta_{LJ}\rangle}{9f^2} \left (\frac{\langle\theta_{LJ}\rangle}{2} \sin\iota_{JN} + \cos\iota_{JN} \right) (\delta f)^2 \\
\cos\theta_{LJ} &\approx \cos\iota_{JN} - \left( \frac{\langle\theta_{LJ}\rangle}{3f} \sin\iota_{JN} \right) \delta f - \frac{\langle\theta_{LJ}\rangle}{9f^2} \left( \frac{\langle\theta_{LJ}\rangle}{2} \cos\iota_{JN} - \sin\iota_{JN} \right) (\delta f)^2 \\
\sin\Phi_{LJ} &\approx 1 - \frac{1}{2} \langle\Omega_{LJ}\rangle^2 \left( \frac{df}{dt} \right)^{-2} (\delta f)^2 \\
\cos\Phi_{LJ} &\approx  -\langle\Omega_{LJ}\rangle \left( \frac{df}{dt} \right)^{-1} \delta f + \frac{\langle\Omega_{LJ}\rangle}{f} \left( \frac{df}{dt} \right)^{-1} (\delta f)^2 \\
\hat{\Vec{L}} \cdot \hat{\Vec{N}} &\approx 1 - \left[ \frac{\langle\theta_{LJ}\rangle^2}{18f^2} + \frac{1}{2} \langle\Omega_{LJ}\rangle^2 \left( \frac{df}{dt} \right)^{-2} \sin^2\iota_{JN} \right](\delta f)^2 \\
\frac{d\delta\Phi}{df} &\approx -\frac{1}{2} \langle\Omega_{LJ}\rangle \left( \frac{df}{dt} \right)^{-1} \frac{\cos\iota_{JN}}{1 + \beta^2} \left[1 + 2\beta^2\left( 1 + \frac{2\tan\iota_{JN}}{\langle\theta_{LJ}\rangle} \right) \right] \label{E:ddeltaPhidf0th}
\end{align}
\end{widetext}
where in Eq.~(\ref{E:ddeltaPhidf0th}) we have defined 
\begin{equation} \label{E:beta}
\beta \equiv \frac{\langle\theta_{LJ}\rangle}{3f\sin\iota_{JN}\langle\Omega_{LJ}\rangle} \frac{df}{dt}
\end{equation}
which is a 2PN order correction.

\section{Validity of evolving the precession parameters according to their lowest PN order scalings
}
\label{Appendix: PN validations}

Our assumption that the dimensionless precession amplitude $\tilde{\theta}$ and frequency $\tilde{\Omega}$ are constant throughout the inspiral implies by Eq.~(\ref{E:PPfreqdep}) that the precession parameters $\langle\theta_{LJ}\rangle$ and $\langle\Omega_{LJ}\rangle$ retain their lowest PN order frequency dependence until merger.
To what degree is this crude approximation valid? To answer this question, we calculate these dimensionless parameters at different frequencies (or binary separations). If the approximation is valid, the dimensionless parameters should be nearly independent of the binary separation. Using Eq.~(\ref{E:PPfreqdep}) and the PN parameter $x \equiv (\pi Mf)^{1/3} = (r/M)^{-1/2}$, we can express the dimensionless parameters as
\begin{subequations}
\begin{align} 
\tilde{\theta} &= 10 \langle \theta_{LJ}  \rangle (4 \eta) \left(\frac{r}{6M}\right)^{1/2}\,, \label{eq: theta_tilde_r} \\
\tilde{\Omega} &= \frac{\langle\Omega_{LJ}\rangle}{10^3\,{\rm Hz}} \left(\frac{M}{M_{\odot}}\right) \left(\frac{r}{6M}\right)^{5/2}, \label{eq: Omega_tilde_r}
\end{align}
\end{subequations}

\begin{figure*}[t!]
    \centering
    \includegraphics[width = 0.49\linewidth]{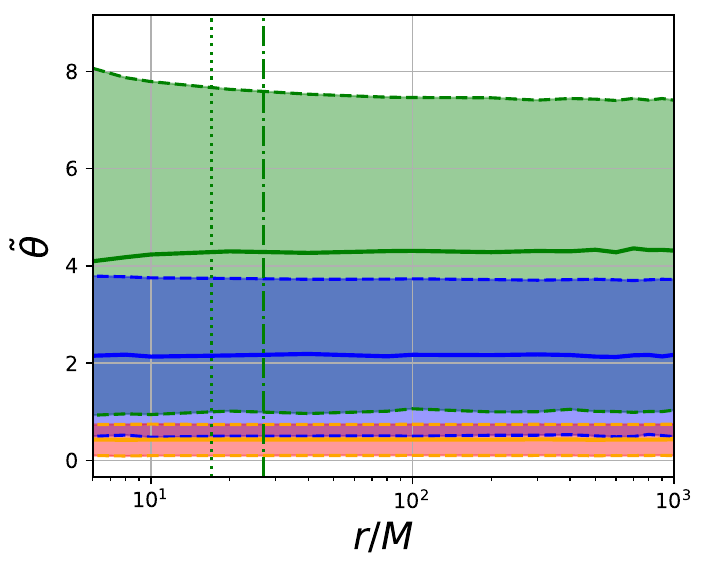}
    \includegraphics[width = 0.49\linewidth]{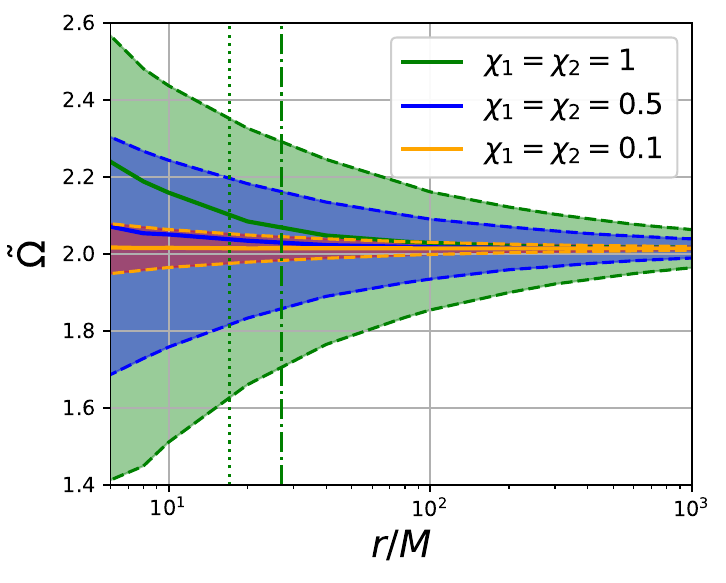}
    \includegraphics[width = 0.49\linewidth]{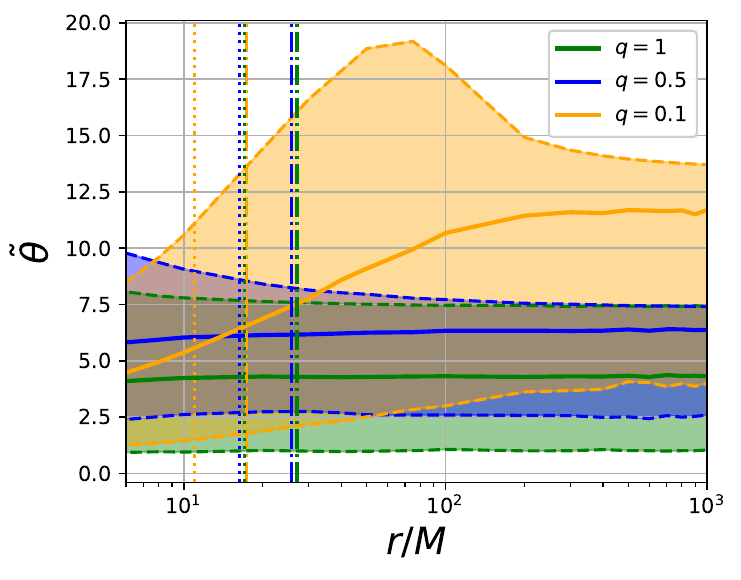}
    \includegraphics[width = 0.49\linewidth]{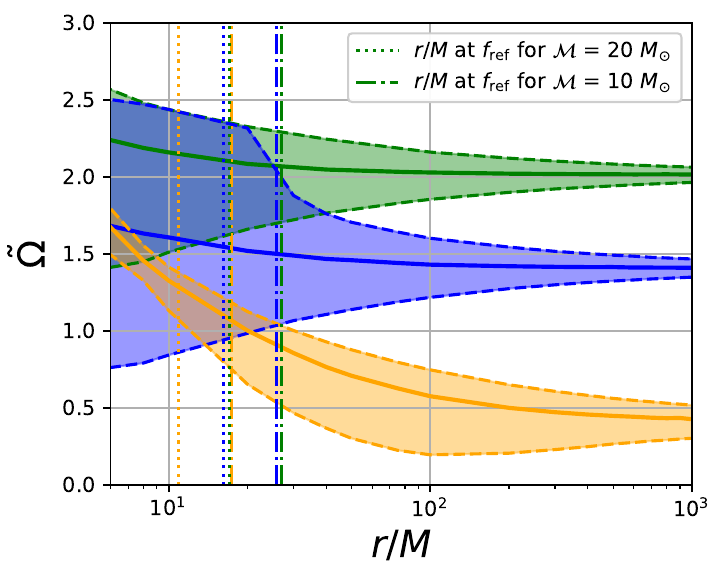}
    \caption{Dimensionless precession parameters $\tilde{\theta}$ and $\tilde{\Omega}$ as functions of binary separation $r/M$ for a population of BBHs with isotropically oriented spins with equal dimensionless magnitudes.  The solid curves show the medians of each distribution, while the lower (upper) dashed curves correspond to the $5^{\rm th}$ ($95^{\rm th}$) percentiles.  The vertical dot-dashed (dotted) lines show the separations at which BBHs with chirp masses $\mathcal{M} = 10 M_{\odot}$ ($20 M_{\odot}$) enter the detector sensitivity band at 20~Hz.  The top panels present equal-mass binaries with dimensionless spins of $\chi_i = 1,\,0.5,$ and $0.1$ shown in green, blue, and orange.  The bottom panels show maximally spinning BBHs with mass ratios of $q = 1,\,0.5,$ and $0.1$ shown in green, blue, and orange.
    }
    \label{fig: PN prec validation}
\end{figure*}

We generate a population of BBHs with isotropically oriented spins of equal dimensionless magnitude ($\chi_1 = \chi_2$), then use the \texttt{precession} package \cite{NewPrecessionCode} to calculate the distributions of the dimensionless precession parameters $\tilde{\theta}$ and $\tilde{\Omega}$ as functions of binary separation from $r = 10^3 M$ to $r = 6M$.  The $5^{\rm th}$, $50^{\rm th}$, and $95^{\rm th}$ percentiles of these distributions are shown in Fig.~\ref{fig: PN prec validation}.

We see that for the equal-mass systems shown in the top panels, the distributions of the dimensionless precession amplitude $\tilde{\theta}$ are barely dependent on binary separation, validating our use of the lowest PN order frequency dependence in Eq.~(\ref{eq: thetaLJ}).  For the dimensionless precession frequency $\tilde{\Omega}$, the median values are nearly constant except for maximal spins for which the median increases by about 10\% throughout the inspiral.  The width of the distributions increase as the BBHs approach merger reflecting the increased importance of spin-spin coupling (not captured in the lowest PN analysis) on the precession frequency.

For BBHs with unequal masses shown in the bottom panels of Fig.~\ref{fig: PN prec validation}, the distributions with $q = 0.5$ scale with binary separation similar to the equal-mass case.  The dimensionless precession parameters are also weakly dependent on binary separation for the smallest mass ratio $q = 0.1$ above $r/M \simeq 100$, but the median amplitude (frequency) seems to decrease (increase) at smaller separations.  This can be understood by recognizing that the primary spin dominates the total angular momentum ($S_1 > L$) for binary separations $r/M < q^{-2}$ ($r/M < 100$ for $q = 0.1$).  At separations below this transition, $\langle \theta_{LJ} \rangle \sim S_i/J$ transitions from $\langle \theta_{LJ} \rangle \sim S_i/L \propto (r/M)^{-1/2}$, i.e. $\tilde{\theta}$ independent of $r$ according to Eq.~(\ref{eq: theta_tilde_r}), to $\langle \theta_{LJ} \rangle \sim S_2/S_1$ independent of $r$, i.e. $\tilde{\theta} \propto (r/M)^{1/2}$.  Below this same transition, $\langle \Omega_{LJ} \rangle$ transitions from the spin-orbit coupling regime $\langle \Omega_{LJ} \rangle \propto (r/M)^{-5/2}$, i.e. $\tilde{\Omega}$ independent of $r$, to the spin-spin coupling regime $\langle \Omega_{LJ} \rangle \propto (r/M)^{-3}$, i.e. $\tilde{\Omega} \propto (r/M)^{-1/2}$ \cite{1995PhRvD..52..821K}.  As we are primarily interested in this work in LVK sources which rarely have $q \lesssim 0.1$, our assumption that the dimensionless precession parameters $\tilde{\theta}$ and $\tilde{\Omega}$ are independent of frequency is reasonably well justified.  This same analysis shows why we chose $\langle \theta_{LJ} \rangle \sim S_i/L \propto \eta^{-1}$ in Eq.~(\ref{eq: thetaLJ}) appropriate for comparable-mass ($q \sim 1$) BBHs.

\section{SNR for distinguishability criterion}
\label{Appendix: Lindblom SNR discussion}

In the Lindblom distinguishability criterion given by Eq.~(\ref{eq: lindblom criterion}), we choose to use the SNR of our source if observed with an ideal template ($h_t = h_s \Longrightarrow \rho^2 = \langle h_s|h_s \rangle$).  We justify this claim below.

Eq.~(5) of \cite{Lindblom} states that two waveforms $h_1$ and $h_2$ are indistinguishable if their difference $\delta h \equiv h_1 - h_2$ satisfies
\begin{align}
    \left< \delta h | \delta h \right> < 1~.
    \label{eq: Lindblom eq5}
\end{align}
Using $\rho_i^2 \equiv \left< h_i | h_i \right>$, this implies
\begin{align}
    {\rho_1}^2 + {\rho_2}^2 - 2 \left< h_1 | h_2 \right> \geq 1~.
    \label{eq: SNR distinguishability}
\end{align}
From the definition of the mismatch (\ref{eq:mismatch}), we get the distinguishability criterion
\begin{align}
    \epsilon \geq \frac{1 - \left(\rho_1 - \rho_2 \right)^2}{2\rho_1 \rho_2}
    \label{eq: symmetric distinguishability}
\end{align}
which differs from the Lindblom criterion (\ref{eq: lindblom criterion}) for $\rho_1 \neq \rho_2$.  We can reconcile these two criteria by setting $1 = s, 2 = t$, and conservatively choosing a template whose luminosity distance $D_t$ maximizes the right-hand side of (\ref{eq: symmetric distinguishability}):

\begin{align}
    \epsilon \geq \max_{D_t}\frac{1 - \left[\rho_s - \rho_{t,0} \left(\frac{D_s}{D_t}\right)\right]^2}{2\rho_s \rho_{t,0}\left(\frac{D_s}{D_t}\right)}
    \label{eq: template distance distinguishability}
\end{align}
where $D_s$ is source luminosity distance, $\rho_{t,0}$ is template SNR evaluated at the source luminosity distance $D_s$, and we have ignored cosmological redshifts.
The right-hand side of (\ref{eq: template distance distinguishability}) is maximized for 
\begin{align}
    \frac{D_s}{D_t} = \frac{\sqrt{{\rho_s}^2 - 1}}{\rho_{t, 0}}
    \label{eq: maximization over D_t}
\end{align}

for which
\begin{align}
    \epsilon \geq 1 - \sqrt{1 - \frac{1}{{\rho_s}^2}}~.
    \label{eq: distinguishability accurate}
\end{align}
In the limit ${\rho_s}^2 \gg 1$, this reduces at lowest order to the Lindblom criterion (\ref{eq: lindblom criterion}).

Template banks of higher dimensionality $D$ will be able to achieve smaller mismatches with a given source.  One can account for this by adopting the stricter distinguishability criterion $\epsilon \geq (D-1)/2{\rho_s}^2$; see Appendix~G of \cite{PhysRevD.95.104004} for more details.  The distinguishability of two families of waveforms can also be assessed in a Bayesian approach; see \cite{Cornish_2015}.

\section{Data release}
\label{Appendix: Data release}

The codes and data supporting the findings reported in this paper are available at the github repository: \url{https://github.com/singhtaman/regular_precession/}. A responsive web app for the waveforms presented in this paper can be found here: \url{https://regular-precession.onrender.com/}. Users can visualize the effects of precession on gravitational-wave signals by changing a binary's intrinsic and extrinsic parameters (along with the precession parameters).

\bibliography{bibme}

\end{document}